%% file: main.tex
\documentclass[11pt,a4paper]{article}

\usepackage[utf8]{inputenc}
\usepackage[T1]{fontenc}
\usepackage{amsmath,amssymb}
\usepackage{booktabs}
\usepackage{graphicx}
\usepackage{siunitx}
\usepackage{xurl}
\usepackage[hidelinks]{hyperref}
\usepackage[margin=2.5cm]{geometry}
\usepackage[numbers,sort&compress]{natbib}
\usepackage{subcaption}
\usepackage{enumitem}
\usepackage{setspace}
\usepackage{lineno}
\ifdefined\submission\linenumbers\fi
\providecommand{\ket}[1]{\left|#1\right\rangle}

\graphicspath{{figures/}}
\title{Trained, Frozen, or Absent:\\
Identifying What a Quantum Layer Contributes to\\
Cross-Sectional Equity Ranking}

\author{Chi-Sheng Chen$^{1,2}$ \quad Xinyu Zhang$^{3}$ \quad En-Jui Kuo$^{4,*}$ \quad
Rong Fu$^{3}$ \quad Qiuzhe Xie$^{5}$ \quad Fan Zhang$^{6}$\\[4pt]
{\small $^{1}$Beth Israel Deaconess Medical Center \& Harvard Medical School, Boston, MA 02115, USA}\\
{\small $^{2}$Brain Research Center, National Tsing Hua University, Hsinchu 300044, Taiwan}\\
{\small $^{3}$Luddy School of Informatics, Computing, and Engineering, Indiana University Bloomington, Bloomington, IN 47405, USA}\\
{\small $^{4}$Department of Electrophysics, National Yang Ming Chiao Tung University, Hsinchu 30010, Taiwan}\\
{\small $^{5}$Institute of Electronics Engineering, National Taiwan University, Taipei 106319, Taiwan}\\
{\small $^{6}$Department of Mathematics, Boise State University, Boise, ID 83702, USA}\\
{\small $^{*}$Corresponding author: kuoenjui@nycu.edu.tw}}
\date{}

\begin{document}
\maketitle

\begin{abstract}
\input{sections/00_abstract}
\end{abstract}


\input{sections/01_introduction}
\input{sections/02_related_work}
\input{sections/03_data}
\input{sections/04_method}
\input{sections/05_protocol}
\input{sections/06_results}

\input{sections/07_discussion}
\input{sections/08_conclusion}

\appendix
\input{sections/A_reproducibility}

\section*{Abbreviations}
\begin{description}[noitemsep,leftmargin=5.5em,labelwidth=5em,font=\normalfont]
  \item[AUC] area under the receiver operating characteristic curve
  \item[B/E] break-even transaction cost
  \item[bps] basis points
  \item[CI] confidence interval
  \item[CNOT] controlled-NOT gate
  \item[DSR] Deflated Sharpe Ratio
  \item[IC] information coefficient
  \item[JPX] Japan Exchange Group
  \item[LSTM] long short-term memory network
  \item[Max DD] maximum drawdown
  \item[MLP] multilayer perceptron
  \item[NYSE] New York Stock Exchange
  \item[OHLCV] open, high, low, close and volume
  \item[PQC] parameterised quantum circuit
  \item[QCNN] quantum convolutional neural network
  \item[QLSTM] quantum long short-term memory network
  \item[QNN] quantum neural network
  \item[QTCNN] quantum temporal convolutional network
  \item[RSI] relative strength index
  \item[RW] Romano--Wolf (stepdown multiple-testing correction)
  \item[SPA] superior predictive ability (test)
  \item[TCN] temporal convolutional network
  \item[TOST] two one-sided tests (of equivalence)
\end{description}

\section*{Declarations}

\subsection*{Availability of data and materials}
The JPX panel is the public competition dataset \citep{jpx2022competition} and
is not redistributed here. The United States panel is reconstructed from public
daily prices by the scripts released with this paper in the GitHub repository
\url{https://github.com/ChiShengChen/qtcnn-trained-frozen-absent}~\citep{chen2026repo}, which also
contain the evaluation stack, the calibration tests, the per-cell results and
daily return series behind every table and figure in this manuscript, the
trained weights of the cells used for the shot-budget measurement, and the
legacy pipelines re-evaluated in Section~\ref{sec:results:reeval}.

\subsection*{Competing interests}
The reference implementation and preprint re-examined in this paper are
the authors' own \citep{chen2025quantumb}. The authors declare that they have no
competing interests.

\subsection*{Funding}
EJK acknowledges financial support from the National Science and Technology
Council (NSTC) of Taiwan under Grant No.~NSTC~114-2112-M-A49-036-MY3. The
funder had no role in the design of the study, the analysis, the decision to
publish or the preparation of the manuscript.

\subsection*{Authors' contributions}
CSC designed the study, wrote the code, ran the experiments and drafted the
manuscript. EJK contributed to the analysis and interpretation of the quantum
diagnostics. XZ, RF, QX and FZ, co-authors of the reference study re-examined
here, contributed to the interpretation of the re-evaluation and to the
discussion. All authors revised the manuscript, and read and approved the
final version.

\subsection*{Acknowledgements}
Not applicable.

\bibliographystyle{unsrtnat}
\bibliography{refs,refs_generated}

\end{document}

%% file: sections/00_abstract.tex
Hybrid quantum--classical models are increasingly proposed for financial
prediction, but reported gains are rarely separated from the classical
components around them. We study a hybrid quantum temporal convolutional
network for daily cross-sectional ranking of \num{2000} Tokyo Stock Exchange
equities under a protocol built to make the quantum contribution
identifiable: six conditions (circuit trained, frozen or absent, a
parameter-matched classical surrogate, and the circuit alone, trained or
frozen) through one code path; four walk-forward folds with five seeds;
bootstrap intervals, Deflated Sharpe Ratios, backtest-overfitting
probabilities and family-wise error control; five classical baselines
including a gradient-boosted tree; and a locked holdout. On \num{500}
out-of-sample days two one-sided tests place the trained-versus-frozen
difference inside \num{0.03} annualised Sharpe ($p = \num{0.0055}$, minimum
detectable effect \num{0.02}): the circuit is equivalent to a frozen one, not
merely indistinguishable from it. With the circuit as the head's only input
the two are equivalent inside \num{0.25}; against the network with no circuit
the design resolves only \num{0.50}. No condition is distinguishable from zero
(best Deflated Sharpe Ratio \num{0.17}). The holdout reproduces the ablation
on a profitable window, and \num{2000} US equities reproduce it where every
condition loses money. A second ansatz family, with data re-uploading, gives
the same ordering, bounded at \num{0.25}. The circuit is trainable and robust
to gate noise, but its shot budget, measured on the trained models' own
inputs, is a property of the trained cell: \num{8192} shots per security
preserve the ranking at the median cell, against \num{256} on random inputs.
Re-scoring our own earlier preprint on this dataset over \num{241} rather
than \num{22} out-of-sample days leaves it no longer significant. Every
analysis decision taken after results were seen is disclosed, every result is
regenerated by the released code, and the per-cell results and trained
weights are released.

%% file: sections/01_introduction.tex
\section{Introduction}
\label{sec:intro}

\paragraph{Motivation.}
Hybrid quantum--classical models are increasingly proposed for financial
prediction, and the reported advantages are typically large: the reference
study this work audits reports a Sharpe ratio \SI{72}{\percent} above its best
classical baseline \citep{chen2025quantumb}. Yet the applied literature
shares three vulnerabilities (short evaluation samples, universes assembled
with hindsight, and weak baselines), and controlled designs that fix
everything but the quantum component find no advantage on the equity
cross-section \citep{shen2026quantum,kobayashi2023cross}. Benchmarks outside
finance reach the same conclusion: prototypical classical baselines beat
prototypical quantum models, entanglement can be removed at no cost
\citep{bowles2024better}, and where a quantum layer appears to help, a
parameter-matched classical control usually helps as much
\citep{gillani2026quantum}. At the same time, the finance literature has
developed tools (the Deflated Sharpe Ratio, the probability of backtest
overfitting, multiplicity-robust inference, decision-time leakage audits)
that the quantum finance literature has not yet adopted; no study in our
literature search reports any of them for a quantum model.

\paragraph{This is a self-audit.}
The reference study whose result this paper re-examines is our own
\citep{chen2025quantumb}. We state that at the outset because it determines how
the rest should be read. The protocol below was not assembled to test someone
else's claim; it was assembled to test a claim we had already made public, and it
returned the opposite answer. Nothing in the analysis depends on that fact:
the conditions and the windows were fixed before the result, the tests that were
revised after it are listed in Section~\ref{sec:protocol:changes}, and the data
and code are released. But a reader is entitled to know
that the paper's critical findings are directed at its own authors' prior work
rather than at a third party's. A literature search of \num{2107} deduplicated
records for this paper found a single study that drew critical conclusions from
its own empirical results \citep{bowles2024better}.

\paragraph{The identification problem.}
A hybrid model typically routes both a quantum score and the classical encoder
representation into a shared head. When such a model outperforms a baseline, the
result is consistent with at least four explanations: the circuit learned a
useful nonlinearity; the circuit acts as a fixed random feature; the classical
encoder did all the work; or the difference is sampling noise. Distinguishing
them requires control conditions, not a larger table of baselines.

\paragraph{Contributions.}
\begin{enumerate}
  \item A controlled ablation that isolates the quantum contribution into six
    conditions (trained circuit, frozen circuit, parameter-matched classical
    surrogate, no circuit, and the circuit alone, trained and frozen) trained
    through an identical code path, with every result regenerated by the
    released code.
  \item An evaluation protocol for cross-sectional ranking that reports
    annualised metrics on \num{500} walk-forward out-of-sample days with
    bootstrap confidence intervals, Deflated Sharpe Ratios, backtest-overfitting
    probabilities and multiple-comparison control.
  \item A batched formulation of the quantum convolutional layer that reproduces
    the per-sample implementation exactly in double precision while running
    \num{33} to \num{72} times faster depending on batch size, which is what
    makes full-sample walk-forward evaluation tractable.
  \item A re-evaluation of our own earlier preprint
    \citep{chen2025quantumb} on \num{241} out-of-sample days rather than the
    \num{22} it reported, which decomposes the advantage it claimed into four
    identifiable sources, one of which is a split adjustment applied in the
    wrong temporal direction that leaves more artefactual returns than
    performing no adjustment at all.
\end{enumerate}

%% file: sections/02_related_work.tex
\section{Related work}
\label{sec:related}

\paragraph{Quantum machine learning for financial prediction.}
Reports of quantum models for forecasting and stock selection range from
clear positives on synthetic data (a 96-parameter circuit matching
thousands-parameter BiLSTMs on noisy sinusoids \citep{emmanoulopoulos2022quantum}) to mixed results on market data: a hybrid network beats a parameter-matched
ANN on directional classification of single stocks, yet its recurrent variant
wins on risk-adjusted returns in only two of four S\&P~500 regimes under
identical walk-forward splits \citep{ahmad2026quantum}; on OHLCV data no
model beats the naive baseline intraday and quantum models merely match
classical ones daily \citep{shah2026stock}; and adding quantum layers to a
Transformer degrades performance on most tasks \citep{chen2025quantumd}. The
two studies closest to ours are both controlled negatives on the equity
cross-section. On TOPIX500 over a ten-year rolling backtest the quantum neural
network is inferior to its classical counterpart over the whole period
\citep{kobayashi2023cross}. On China A-shares, a kernel-swap design that holds
subsamples, solver and tuning budget fixed finds the fidelity quantum kernel
indistinguishable from an RBF control ($\Delta\mathrm{IC}=+0.005$, $p=0.42$)
over 170 walk-forward windows, with no pairwise difference among eleven models
surviving family-wise correction, and shows that a 60-window evaluation on a
hindsight-screened universe makes the same kernel look dominant
\citep{shen2026quantum}. That paper's diagnosis of the applied literature
(short samples, hindsight universes, weak baselines) and its proposed
standards (kernel-swap controls, budget-equalised comparisons, point-in-time
universes, multiplicity-robust inference) are the ones our protocol adopts. The
reference QTCNN study we audit \citep{chen2025quantumb} reports an annualised
out-of-sample Sharpe ratio of \num{0.538} $\pm$ \num{0.042} over a 22-day
window, acknowledges that 22 days give limited power, and defers a full-sample
backtest to future work; its own encoder-only ablation scores \num{0.412},
which already places most of the reported edge in the classical encoder.

\paragraph{QCNN trainability and dequantisation.}
Randomly initialised parametrised circuits generically exhibit barren
plateaus \citep{mcclean2018barren,larocca2024barren}; global cost functions
induce them even at shallow depth whereas local observables keep gradients
polynomial up to depth $O(\log n)$ \citep{cerezo2021cost}, hardware noise
induces its own plateau \citep{wang2021noise}, and data embeddings add a
dataset-induced variant \citep{thanasilp2021subtleties}. The QCNN of
\citet{cong2019quantum} avoids the problem: its gradient variance vanishes at
most polynomially, and sharing parameters across a layer, as in the original
design, enlarges the variance further \citep{pesah2020absence}. Our
gradient-variance and expressibility scans (Section~\ref{sec:results}) are the
empirical counterpart of these guarantees for the ansatz we use. The guarantee
cuts both ways. Randomly initialised QCNNs act only on low-weight information
about their inputs, and a classical surrogate built from Pauli propagation
matches or beats them on every benchmark up to 1024 qubits
\citep{bermejo2024quantum}; more generally, the structure that removes barren
plateaus tends to confine the model to a classically simulable subspace
\citep{cerezo2023provable}. Classical surrogates that reproduce a trained
circuit's input--output map are therefore the natural benchmark a quantum model
must beat \citep{schreiber2022classical}; random-Fourier-feature surrogates can
be built from the architecture alone \citep{landman2022classically}; wide
random circuits converge to Gaussian processes and stay there under training
\citep{girardi2024trained}; and in the kernel view the optimal model is a convex
problem \citep{schuld2021quantum}. Our \textsc{Frozen} and \textsc{Surrogate}
conditions operationalise this random-feature reading: if the trained circuit
is no better than a frozen one, and the frozen one no better than a classical
map of equal parameter count, the layer is a random feature and the
dequantisation argument applies without further analysis. Related diagnostics
warn against reading performance as evidence of quantum effects: hybrid
networks show no consistent expressibility--trainability relation
\citep{kashif2026rethinking}, a fixed encoding can impose a capacity ceiling
that mimics a plateau \citep{pandey2026architectural}, current models fit random
labels so uniform generalisation bounds are vacuous
\citep{gilfuster2023understanding}, and a causal-mediation analysis finds
quantum-mediated effects averaging \SI{0.82}{\percent} with \SI{93}{\percent} of
configurations in a neutral regime \citep{kesiku2026quantum}.

\paragraph{Benchmarking practice in hybrid quantum ML.}
Systematic benchmarks find prototypical classical baselines beating
prototypical quantum models, with entanglement removable at no cost and only
one paper in the authors' literature sample drawing critical conclusions from
its own results \citep{bowles2024better}; an industrial comparison finds
classical and hybrid models statistically equivalent, with accuracy
uncorrelated with entangling power \citep{basilewitsch2024quantum}. Budget-aware
protocols that match trainable parameters and optimisation steps find shallow
hybrids matching classical performance at much higher cost
\citep{bendouma2026fair}, whereas a same-accuracy-threshold comparison reports
hybrids scaling FLOPs more efficiently \citep{kashif2024computational}. The
attribution audit of \citet{gillani2026quantum} (parameter-matched classical
controls sharing the identical front-end, a random-feature kernel, a
regularisation sweep, three to five seeds with confidence intervals, and the
explicit stance that the benchmark's value does not depend on quantum winning) is the closest methodological template to ours; the capacity-matched
bottleneck of \citet{chen2025quantumd} follows the same commitment. Shot
budgets are the operative hardware cost, since access is priced per shot
\citep{liu2025you} and gradient estimation needs two circuit runs per parameter
per iteration \citep{scriva2023challenges,miroszewski2024search}; batched
parameter broadcasting is what removes per-call simulator overhead
\citep{chen2025batched}.

\paragraph{Evaluation methodology in empirical finance.}
A handful of trials suffices to find a spuriously high backtested Sharpe ratio
\citep{bailey2014pseudo}, calibrating rules by backtest itself overfits
\citep{carr2014determining}, and standard holdout is unreliable, motivating the
Deflated Sharpe Ratio \citep{bailey2014deflated} and the combinatorially
symmetric estimate of the probability of backtest overfitting
\citep{bailey2017probability} that we report. Adaptive specification search
produces significant backtests under martingale-difference nulls, and effective
multiplicity rather than the nominal specification count sets the expected
in-sample maximum \citep{nikolopoulos2026spurious}; publication selects for
author and sampling error alike \citep{chen2022publication}. Multiple-testing
control in the cross-section is contested in magnitude (93\% of
trading-friction anomalies fail replication and microcaps drive equal-weighted
results in \citet{hou2020replicating}, while a Bayesian replication finds the
majority of factors survive \citep{jensen2021there}) but not in principle,
which is why we apply Romano--Wolf step-down control \citep{romano2005stepwise}
across every contrast. Leakage inflates out-of-sample performance while
preserving the appearance of rigour \citep{bhand2026illusion}; a one-switch
benchmark shows chronological splitting alone is insufficient and fixes
decision-time semantics as features known by the close of day~$t$ and execution
at the $t{+}1$ open \citep{zhang2026alpha}, which is the convention our
target definition and embargo follow. Transaction costs are decisive at daily
turnover: technical rules that pass an SPA test in sample do not survive
moderate costs \citep{rink2023predictive}, 73.3\% directional accuracy
coexists with a post-cost loss at 471 trades per year \citep{pall2026transaction},
and a systematic falsification of fourteen intraday signal families finds none
passing a $t\ge 2$, $\ge 30$-trade, net-positive standard
\citep{mesfin2026structural}. Our cost curves and break-even analysis are the
corresponding checks for a cross-sectional long--short book.

%% file: sections/03_data.tex
\section{Data}
\label{sec:data}

We use the daily price panel of the JPX Tokyo Stock Exchange Prediction
competition~\citep{jpx2022competition}, published by the Japan Exchange Group
(JPX), covering \num{2000} securities. The training file spans 2017-01-04 to 2021-12-03 and the
supplemental file continues to 2022-06-24; after feature construction the
usable panel holds \num{2483470} security-days over \num{1296} trading dates
from 2017-02-02. The \num{2000} securities are the competition's primary
universe, a membership list fixed by the organiser when the data were released
at the end of the training period and applied to the whole window. It therefore
carries the hindsight that Section~\ref{sec:intro} criticises: a name is in the
panel for 2017 because it was listed and liquid in late 2021. The bias is common
to every condition and to the earlier result we re-score, so it enters no
contrast between them, and no absolute level of performance on this panel is
claimed.

\subsection{Split adjustment}
\label{sec:data:adjustment}
The panel records a per-day \texttt{AdjustmentFactor}. A factor of $0.5$ on day
$t$ indicates that prices up to and including $t$ must be halved to be
comparable with prices after $t$, so the cumulative factor at date $t$ is the
product of all factors from $t$ onwards:
\begin{equation}
  \tilde{P}_t = P_t \prod_{s \ge t} a_s .
\end{equation}
Publicly circulated pipelines for this dataset instead accumulate forward,
$\prod_{s \le t} a_s$. Table~\ref{tab:adjustment} shows the consequence over the
\num{610} securities with at least one non-unit factor.

\begin{table}[t]
\centering
\caption{Artefactual daily moves under three adjustment conventions.}
\label{tab:adjustment}
\begin{tabular}{lrrr}
\toprule
Adjustment & $|r|>30\%$ & $|r|>50\%$ & $\max |r|$ \\
\midrule
None (raw closes)            & 718  & 565  & 20.09 \\
Forward cumulative product   & 1418 & 1091 & 23.43 \\
Backward cumulative product  & \textbf{16} & \textbf{3} & \textbf{1.12} \\
\bottomrule
\end{tabular}
\end{table}

\subsection{Features}
Twenty-one per-security features are computed from the adjusted close $\tilde P_t$,
the daily return $r_t=\tilde P_t/\tilde P_{t-1}-1$, the raw volume $V_t$ and the
dollar volume $D_t=P_tV_t$ (Table~\ref{tab:features}). Every window looks
strictly backwards, so a feature dated $t$ uses closes up to and including $t$
only. The set is deliberately generic (momentum, realised volatility,
standardised returns, liquidity, range position, drawdown and RSI) so that
no model has access to information the others lack.

\begin{table}[t]
  \centering
  \caption{The 21 input features. All rolling statistics use a window of
  exactly $n$ trading days and are undefined until $n$ observations exist.}
  \label{tab:features}
  \small
  \begin{tabular}{llc}
    \toprule
    Group & Definition & $n$ \\
    \midrule
    Momentum ($\mathrm{mom}_n$) & $\tilde P_t/\tilde P_{t-n}-1$ & 1, 2, 5, 10, 20 \\
    Volatility ($\mathrm{vol}_n$) & rolling standard deviation of $r$ & 5, 10, 20 \\
    Standardised return ($\mathrm{ret\_z}_n$) & $(r_t-\bar r_n)/\sigma_n(r)$ & 5, 10, 20 \\
    Log volume & $\log(1+V_t)$ & -- \\
    Log dollar volume & $\log(1+D_t)$ & -- \\
    Dollar-volume $z$-score ($\mathrm{dv\_z}_n$) & rolling $z$-score of $\log(1+D_t)$ & 5, 10, 20 \\
    Range position ($\mathrm{pos}_n$) & $(\tilde P_t-\min_n)/(\max_n-\min_n)$ & 10, 20 \\
    Drawdown ($\mathrm{dd}_{20}$) & $\tilde P_t/\max_{20}-1$ & 20 \\
    RSI ($\mathrm{rsi}_n$) & simple-moving-average RSI, scaled to $[-0.5,0.5]$ & 7, 14 \\
    \bottomrule
  \end{tabular}
\end{table}

\subsection{Cross-sectional normalisation}
Each feature is winsorised at the 1st and 99th percentile and standardised
within each trading date. Both operations use only same-day information and
therefore introduce no look-ahead.

\subsection{Second market}
\label{sec:data-us}
To test whether the result is a property of one exchange, the identical
pipeline is run on United States common stocks over the same calendar window.
Candidates are the \num{4826} NYSE, NYSE American and Nasdaq common stocks in
the exchange symbol directories after removing ETFs, warrants, units, rights,
preferred shares, notes, depositary receipts, trusts and acquisition shells by
name; \num{3935} of them traded in the window. The universe is the \num{2000}
names with the highest median dollar volume in 2017, the first calendar year,
among those with at least \num{200} trading days and a median close of at least
\$2 that year, so selection uses no information after the first training
window. Split-adjusted closes and volumes come from a public daily-price source and
are adjusted by the convention of Section~\ref{sec:data:adjustment}; the
target, the 21 features and the
cross-sectional normalisation are computed by the same functions. A
point-in-time price floor drops security-days whose close is below \$1, the
standard penny-stock screen. It removes \num{4670} security-days (\SI{0.17}{\percent})
across 49 symbols, among them a bankruptcy-reorganisation splice in which the
price source joined the post-emergence entity to the pre-emergence ticker and
produced a single forward return of \SI{25733}{\percent}; that one
security-day contaminated our first run on this panel
(Section~\ref{sec:results:us}). The panel then holds \num{2792249} security-days over \num{1399} trading dates from
2016-11-30 to 2022-06-22. Because the US calendar has more trading days, the
development period yields five walk-forward folds (\num{625} out-of-sample
days) rather than four; the holdout window is the same. The symbol directory
is a current snapshot, so names delisted before it are absent; every condition
is trained and scored on this identical panel, so the survivorship bias is
common to all conditions and does not enter the contrasts between them,
and absolute performance levels on this market are not claimed.

%% file: sections/04_method.tex
\section{Model}
\label{sec:method}

\subsection{Architecture}
\label{sec:method:arch}
The model maps a window of $T=20$ daily cross-sectionally normalised feature
vectors $\mathbf{x}_{1:T} \in \mathbb{R}^{T \times F}$, $F=21$, to a scalar
ranking score in three stages.

\paragraph{Temporal encoder.}
A dilated causal convolutional stack reduces the window to a vector
$\mathbf{e} \in \mathbb{R}^{n}$, where $n$ is the qubit count. Two blocks with
kernel size 3 and dilations $1$ and $2$ give a receptive field of seven trading
days; the representation of the final step is projected through a two-layer
head whose $\tanh$ output is scaled by $\pi$, so that every component is a valid
rotation angle. Padding is applied on the left only. A symmetric padding, as
used in the reference implementation this work audits, mixes each position with
the step that follows it and so does not implement the receptive field the
architecture claims.

\paragraph{Quantum convolutional layer.}
The encoder output is embedded as single-qubit rotations,
$\ket{\psi_0} = \bigotimes_{i=1}^{n} R_Y(e_i)\ket{0}$, and processed by
$L = \min(L_{\max}, \lfloor \log_2 n \rfloor)$ convolution-and-pooling stages.
Each stage applies the two-qubit block
\begin{equation}
  U(\boldsymbol{\theta}) =
  \bigl(R_Y(\theta_5) \otimes R_Y(\theta_6)\bigr)\,
  \mathrm{CNOT}_{21}\,
  \bigl(R_Z(\theta_3) \otimes R_Z(\theta_4)\bigr)\,
  \mathrm{CNOT}_{12}\,
  \bigl(R_Y(\theta_1) \otimes R_Y(\theta_2)\bigr)
\end{equation}
to disjoint neighbouring pairs, then pools by retaining every second wire. The
same $\boldsymbol{\theta} \in \mathbb{R}^{6}$ is reused across all pairs within
a stage, so the circuit carries $6L$ parameters in total (18 at $n=8$,
$L=3$) independently of the register width. The layer output is the
expectation $\langle Z \rangle$ of the single surviving wire.

\paragraph{Second ansatz family.}
To test whether the result depends on the circuit family, the
trained-versus-frozen contrast is repeated on a circuit that differs from the
QCNN in every structural respect: the encoder output is
re-uploaded as $R_Y$ rotations before each of $L = 3$ layers; each layer
applies $R_Y(\theta)\,R_Z(\phi)$ to every wire, with its own parameters, and
entangles the register with a ring of CNOTs; there is no pooling; and the
output is the mean of $\langle Z \rangle$ over all eight wires. It carries
$2nL = 48$ parameters, re-uploading raises the Fourier degree of the function
it computes, and the readout is a sum of single-qubit $Z$ terms over every
wire rather than one wire's $Z$ (a sum of local observables, not a many-body
operator, so the barren-plateau results for global cost functions do not apply
to it). It is run in the
\textsc{Trained} and \textsc{Frozen} conditions on the Tokyo panel, with the
same seeds, folds, encoder, head and training loop, and compared with the
same \textsc{None} control.

\paragraph{Head.}
A two-hidden-layer perceptron consumes the concatenation
$[\langle Z \rangle, \mathbf{e}]$ and emits a logit. Training minimises binary
cross-entropy against cross-sectional extreme labels
(Section~\ref{sec:protocol:labels}).

\subsection{Batched circuit evaluation}
\label{sec:method:batched}
Evaluating the QNode once per sample, as the reference implementation does, is
the dominant cost of training. Passing the batch through the simulator as a
broadcast parameter dimension yields bit-identical expectation values (the
maximum absolute difference over every configuration we timed is exactly zero
in double precision) while replacing the Python-level loop. At $n=8$ and
$L=3$ the forward pass over a batch of \num{512} falls from
\SI{8.64}{\second} to \SI{0.12}{\second}, a factor of \num{71.7}; the gain
grows with batch size, from \num{33.1}$\times$ at \num{64} through
\num{64.2}$\times$ at \num{256}, because the fixed per-call overhead is
amortised over more samples (Table~\ref{tab:batching}). Adjoint differentiation
on a state-vector backend is slower at this width because the register is too
small to amortise its own overhead.

\begin{table}[t]
  \centering
  \caption{Batched against per-sample circuit evaluation at $n=8$, $L=3$,
  \num{18} parameters. Median of three repeats after warm-up; the machine was
  under scheduler control. Outputs agree exactly in double precision.}
  \label{tab:batching}
  \begin{tabular}{rrrrr}
    \toprule
    Batch & Per-sample loop (s) & Batched (s) & Speed-up & $\max|\Delta|$ \\
    \midrule
    64  & 0.765 & 0.023 & 33.1$\times$ & 0 \\
    256 & 3.218 & 0.050 & 64.2$\times$ & 0 \\
    512 & 8.635 & 0.120 & 71.7$\times$ & 0 \\
    \bottomrule
  \end{tabular}
\end{table}

The speed-up matters for the evaluation. The reference implementation compensated
for its cost by subsampling to 9\% of trading days and by capping training at
\num{10000} sequences, which is what reduced its reported out-of-sample windows
to 22 and in some runs 2 trading days. The speed-up is what makes the evaluation
protocol of Section~\ref{sec:protocol} affordable.

\subsection{Ablation conditions}
\label{sec:method:ablation}
A hybrid model whose head also observes the classical encoding cannot, on its
own, tell us whether the circuit contributed. We therefore train four conditions
through an identical code path, differing only in the block that occupies the
quantum position:

\begin{description}
  \item[\textsc{Trained}] the circuit parameters $\boldsymbol{\theta}$ are
    learned jointly with the encoder and head.
  \item[\textsc{Frozen}] the identical circuit, held at its initialisation,
    $\boldsymbol{\theta} \sim 0.1\,\mathcal{N}(0, I)$, the same draw per seed
    from which \textsc{Trained} starts. The contrast against \textsc{Trained} separates the value of
    \emph{optimising} the circuit from the value of the fixed nonlinear feature
    it computes.
  \item[\textsc{Surrogate}] a classical map with the same parameter count and
    the same pairwise-mixing-then-pooling structure, with angles initialised on
    $(-\pi, \pi]$ so that its output distribution matches the circuit's in
    scale (mean absolute activation \num{0.59} against \num{0.69}). A
    small-normal initialisation, the obvious default, shrinks its output by
    three orders of magnitude across three layers and would handicap the
    control.
  \item[\textsc{None}] the head receives $\mathbf{e}$ alone.
\end{description}

A fifth and a sixth condition, \textsc{QuantumOnly} with the circuit trained
and with it frozen, withhold $\mathbf{e}$ from the head so that the score
depends on the circuit output alone. The pair bounds how much ranking
information the measured observable carries by itself, and it is the one
contrast in which the circuit's parameters are the only route from features to
score, so trained-versus-frozen there cannot be dismissed as structural
(Section~\ref{sec:results:ablation}).

Parameter counts across the four hybrid conditions differ by less than
0.4\% (\numrange{23817}{23899}); the two circuit-only conditions, whose head
has eight fewer inputs, carry \num{23387} and \num{23369}. The classical
baselines below are sized as stand-alone models, not matched to these.

\paragraph{Classical baselines.}
Five models somebody would deploy, trained on the same windows, labels,
subsample and portfolio rule: a wider temporal CNN (hidden width \num{128},
three blocks), an LSTM, a Transformer encoder, a perceptron on the flattened
window, and a gradient-boosted tree (LightGBM, \num{63} leaves, learning rate
\num{0.03}, up to \num{2000} trees with early stopping on validation AUC after
\num{100} rounds) on the same flattened $20 \times 21$ window. The last is the
model family that wins cross-sectional equity competitions, including the one
this dataset comes from, so its absence would be the first thing a finance
reader noticed.

%% file: sections/05_protocol.tex
\section{Evaluation protocol}
\label{sec:protocol}

\subsection{Walk-forward folds}
\label{sec:protocol:folds}
The development period, 2017-02-02 to 2021-12-03, holds \num{1161} trading
days and is divided into four rolling-origin folds. Each fold trains on
\num{500} trading days, selects on the following \num{60}, discards an embargo
of \num{20}, and tests on the next \num{125}. A fifth fold would need
\num{1205} days and is therefore not available. The embargo matters because features use rolling windows of up to
\num{20} days: without it, the input windows of the first test days overlap the
last training days. Test blocks are disjoint by construction, so their return
series concatenate into a single continuous out-of-sample track of \num{500}
trading days.

Rolling rather than anchored windows keep the training sample size constant
across folds, which separates regime effects from sample-size effects.

\subsection{Labels}
\label{sec:protocol:labels}
Within each training date the \num{200} securities with the lowest realised
forward return are labelled $0$ and the \num{200} highest are labelled $1$. The
task is ranking rather than level prediction, and the middle of the
cross-section carries almost no separable signal, so training on the tails gives
a cleaner decision boundary than regression on all names.

\subsection{Training}
\label{sec:protocol:training}
Every condition is trained by the same loop: AdamW (learning rate $10^{-3}$,
weight decay $10^{-4}$) under a cosine schedule over a budget of \num{30}
epochs, batches of \num{512}, gradient norm clipped at \num{1}, binary
cross-entropy on the extreme labels, and a random subsample of \num{60000}
labelled sequences per fold drawn by the seed. Early stopping monitors the
validation AUC with a patience of six epochs and restores the best weights; on
the walk-forward folds it halts after \num{9.1} epochs on average for the
circuit conditions. Because that could be read as never giving the circuit time
to matter, both circuit conditions are also run for the full \num{30} epochs
and scored on the last epoch's weights (Section~\ref{sec:results:ablation}).

\subsection{Portfolio construction}
Model outputs are standardised within each trading date, restricted to names
that are in the primary universe, not under supervision, and have positive close
and volume, and the top and bottom \num{200} are held long and short with equal
weights. The daily strategy return is the spread between the two legs.

\subsection{Statistics}
\label{sec:protocol:stats}
All Sharpe and Sortino ratios are annualised by $\sqrt{252}$. Information
coefficients are the within-date Spearman correlations between score and
realised forward return; because that series is autocorrelated, its mean is
tested with Newey--West standard errors at the automatic lag
$4(n/100)^{2/9}$~\citep{newey1987simple,newey1994automatic}. Confidence intervals for the Sharpe
ratio come from a stationary bootstrap with mean block length
\num{10}~\citep{politis1994stationary}, which brackets the automatic selection
of \citet{politis2004automatic} on these series; the null is imposed on the
return series by demeaning it before resampling, rather than by re-centring the
bootstrap draws of the statistic, because the latter assumes a symmetric
sampling distribution that a skewed return series does not have. Intervals
are percentiles of the ordinary bootstrap distribution and $p$-values come from
the null-imposed one, so near the boundary the two can disagree: an interval
can just exclude zero while $p$ stays above \num{0.05}. The $p$-value is the
test, and each such case is pointed out where it occurs.

A non-significant difference is not evidence of no difference, so every null
claim below is accompanied by the two statistics that make it interpretable.
The minimum detectable effect is the smallest true difference the design would
have found at \SI{80}{\percent} power given its standard error. Equivalence is
assessed by two one-sided tests against a stated
margin~\citep{schuirmann1987comparison}, applied to the paired bootstrap
distribution of the Sharpe difference; a significant result there is positive
evidence that the difference lies inside the margin. The margin of \num{0.03}
annualised Sharpe used for the trained-versus-frozen contrast was not fixed in
advance. The equivalence analysis entered the protocol after the walk-forward
results were in, when the pooled-track Sharpe correlation we had first reported
as the headline turned out to be location-invariant and therefore uninformative,
and \num{0.03} was chosen at that point as the smallest round margin the paired
standard error (\num{0.007}, minimum detectable effect \num{0.02}) could
resolve. A margin fixed with the standard error in view is a weaker commitment
than one fixed before the data, so the difference and its standard error are
always printed beside the verdict, letting the reader apply any margin, and the
contrasts the design resolves less finely are reported at the wider margins they
require (\num{0.25} and \num{0.50}) rather than at one the data cannot support.
Section~\ref{sec:protocol:changes} lists every change made to the analysis after
results were seen.

Two further statistics guard against the selection bias that a study comparing
several model variants necessarily incurs. The Deflated Sharpe
Ratio~\citep{bailey2014deflated} discounts the observed Sharpe by the expected
maximum across the number of configurations searched, which we report
explicitly. The probability of backtest overfitting~\citep{bailey2017probability}
is estimated by combinatorially symmetric cross-validation over the candidate
set; our implementation returns \num{0.53(4)} on independent random strategies,
consistent with the uninformative value of \num{0.5}.

Pairwise contrasts use the Diebold--Mariano test on the difference of daily
return series~\citep{diebold1995comparing}, with the small-sample correction of
\citet{harvey1997testing} because a window of a few hundred days is short for
the normal approximation. That statistic tests a difference in mean return,
which is not the hypothesis the tables are ordered on, so each contrast is also
tested directly on the difference of Sharpe ratios by a paired stationary
bootstrap that resamples both tracks on the same days.

Family-wise error is controlled by the Romano--Wolf stepdown
procedure~\citep{romano2005stepwise}, which is far less conservative than
Bonferroni at this number of comparisons. We state the family explicitly: it is
the set of contrasts against the no-circuit control on one panel: ten
hypotheses on the Tokyo walk-forward, eight on the US panel and eight on the
holdout. It does not cover the tests of each condition against zero, and it is
not pooled across the two markets or the holdout; those are reported
uncorrected and should be read as such.

\subsection{Quantile portfolios}
\label{sec:protocol:quantiles}
The information coefficient is a rank correlation over the whole
cross-section, whereas the long/short book holds only the two tails. The two
can therefore disagree, and on this dataset they do: in the reference
implementation the model with the lowest IC has the highest Sharpe
(Section~\ref{sec:results}). We therefore report decile portfolios: the mean
forward return of each score decile with Newey--West $t$-statistics, the
top-minus-bottom spread, and the Spearman correlation between decile index and
mean return as a monotonicity measure. A signal with a high IC but low
monotonicity does not belong in a tail-based long/short book, and saying so is
more useful than reporting either statistic alone.

\subsection{Transaction costs}
\label{sec:protocol:costs}
Turnover is the fraction of each leg replaced between consecutive rebalances.
Net returns charge a round-trip cost on the traded fraction of both legs, and we
report Sharpe ratios on a grid of \numlist{0;5;10;20;30;50} basis points
together with the break-even cost at which the mean net return reaches zero.
Reporting gross performance alone is not meaningful for a daily-rebalanced
book: on the reference implementation's own output, turnover averages 24\% per
day and the break-even cost is \SI{7.4}{bps}.

\subsection{Locked holdout}
\label{sec:protocol:holdout}
The supplemental window, 2021-12-06 to 2022-06-24 (\num{135} trading days), is
evaluated once per condition, after that condition's architectural and
hyperparameter decisions are frozen. It was opened three times: in September
2026 for the eight network conditions; in October 2026 for the LightGBM
baseline, added to the study after the first opening; and in November 2026,
when the eight network conditions were re-run on it, unchanged, under the
released code, because the September run had been made with an uncommitted
state of the code and could not be reproduced
(Section~\ref{sec:protocol:changes}). No condition's design was altered at any
opening. The tables report the third opening, which anyone can reproduce, and
Appendix~\ref{sec:app:rerun} reports the first beside it. The circuit-only
conditions and the second ansatz family were never run on the window, which
spans a market regime absent from the development period.

\subsection{Changes made after results were seen}
\label{sec:protocol:changes}
A protocol section that reads as if everything had been fixed in advance would
overstate this study's discipline. The following were changed after
walk-forward results existed, and each is reported in full rather than
replaced.
\begin{enumerate}
  \item The trained-versus-frozen headline was changed from a pooled-track
    Sharpe correlation to the equivalence test above, and the \num{0.03} margin
    was chosen with the standard error in view (September 2026).
  \item Four inference defects were corrected: the bootstrap null is now imposed
    on demeaned returns rather than by re-centring Sharpe draws; Sharpe
    differences are tested as such by a paired bootstrap, not only through the
    Diebold--Mariano test on mean returns; the Harvey--Leybourne--Newbold
    small-sample correction and a $t$ reference are used for Diebold--Mariano;
    and the Romano--Wolf family is stated explicitly. At the time, the only
    verdict that changed was that three US conditions became significantly
    negative.
  \item The dispersion term in the Deflated Sharpe Ratio was changed from the
    sampling standard error $1/\sqrt{n}$ to the empirical standard deviation of
    the \num{55} trial Sharpe ratios, which is the quantity the expected-maximum
    formula of \citet{bailey2014deflated} takes. Every DSR in this paper uses the
    latter; on the first run the former gave far smaller values (\num{0.012}
    rather than \num{0.17} for the best walk-forward condition) and the same
    verdict.
  \item The shot budget was re-measured at the input distribution the deployed
    model produces, and then on the trained encoders rather than on an
    initialised one (Section~\ref{sec:results:diagnostics}).
  \item Conditions were added after the first results: the circuit-only pair,
    the no-early-stopping control, the second ansatz family and the LightGBM
    baseline. All are reported; none was removed. The Deflated Sharpe Ratio
    counts the \num{55} configurations of the main comparison (eleven
    conditions, five seeds). The second-ansatz and full-budget controls add
    \num{20} further configurations on the same track, each testing a single
    contrast; they are not counted, and counting them would lower every
    Deflated Sharpe Ratio further.
  \item Every network result in the paper was regenerated in November 2026
    with the released code: both walk-forward panels, the circuit-only
    condition, the full-budget control and the holdout. The September runs had
    been made from working copies of the code that were never committed. The
    committed code is
    deterministic, in that repeated runs of a cell are bit-for-bit identical at
    one and at four threads, yet it reproduced none of the September cells;
    on the two walk-forward panels, per-cell Sharpe ratios correlate at
    \numrange{0.40}{0.82} between the two runs of the same condition, while the trained and frozen circuits within
    either run still agree to \num{0.999}. Appendix~\ref{sec:app:rerun}
    compares the two runs condition by condition. Every conclusion survived;
    the numbers in this paper are the reproducible ones.
  \item The full-budget control (Section~\ref{sec:results:ablation}) trained
    for 30 epochs but, in the released code, then restored the
    best-validation weights, which made it bit-for-bit identical to the
    early-stopped run. It now scores the weights of the last epoch, as the
    September run had.
  \item The locked holdout was opened a second time for the LightGBM baseline
    and a third time for the reproducible rerun
    (Section~\ref{sec:protocol:holdout}).
\end{enumerate}

%% file: sections/06_results.tex
\section{Results}
\label{sec:results}

All figures in this section are computed on the same \num{500} out-of-sample
trading days (2019-06-18 to 2021-08-05), pooled from four disjoint walk-forward
test blocks, with each condition averaged over five seeds. Seed-averaging a
date's return is the return of an equal-weight ensemble of the seeds, which is
what a practitioner would hold; it removes optimisation noise without treating
the folds as overlapping.

\subsection{Does the circuit contribute?}
\label{sec:results:ablation}

\begin{figure}[t]
\centering
\includegraphics[width=\linewidth]{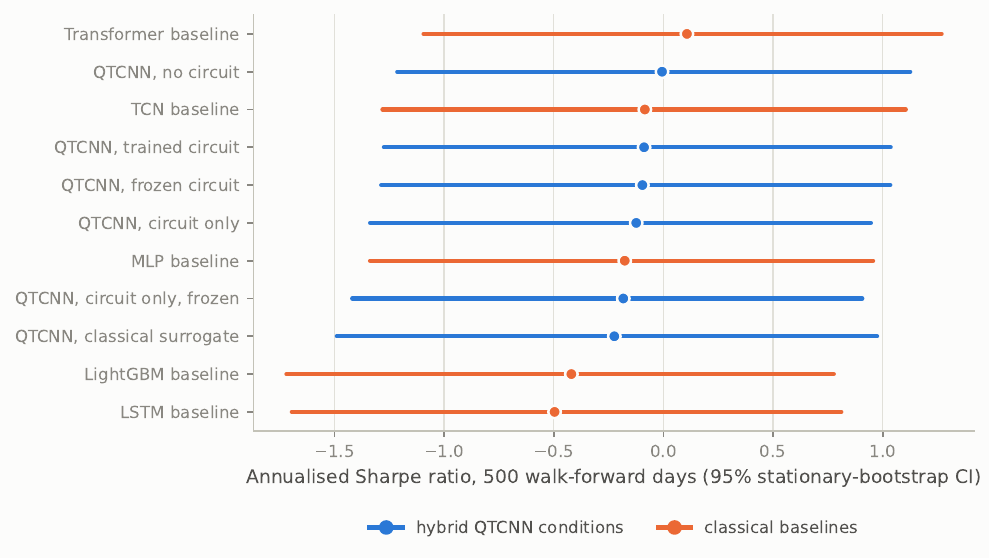}
\caption{Annualised Sharpe ratio of each condition on the pooled 500-day
walk-forward track, with 95\% stationary-bootstrap confidence intervals. The
four hybrid conditions differ only in what occupies the quantum position.}
\label{fig:ablation}
\end{figure}

Figure~\ref{fig:ablation} and Table~\ref{tab:headline} give the headline. No
condition is distinguishable from zero: every 95\% stationary-bootstrap
interval spans it, the bootstrap $p$-values run from \num{0.44} to \num{0.99},
and the best Deflated Sharpe Ratio, deflated by the \num{55} configurations
(eleven models, five seeds) the study evaluated, is \num{0.17} against the
\num{0.95} that a claim would require.

\input{tables/headline}

Nor is any condition distinguishable from the \textsc{None} control, the same
network with the circuit removed. Diebold--Mariano tests on the aligned daily
return series give $p = 0.53$ for \textsc{Trained}, \num{0.49} for
\textsc{Frozen}, \num{0.09} for \textsc{Surrogate} and \num{0.80} for the TCN
baseline; after Romano--Wolf stepdown correction the smallest adjusted $p$ in
the family of ten is \num{0.38}, for the surrogate.

The circuit-only pair bounds the question from the other side. \textsc{QuantumOnly},
whose head sees the circuit output and nothing else, reaches a pooled Sharpe
ratio of \num{-0.124} (95\% CI $[-1.34, +0.95]$, $p = 0.82$) and an information
coefficient of \num{0.0075} ($t = 0.91$), against \num{0.0082} for the
no-circuit control (Table~\ref{tab:ic}). It sits \num{0.12} Sharpe below that
control, with a paired standard error of \num{0.16} ($p = 0.43$;
Diebold--Mariano $p = 0.49$, Romano--Wolf $p = 0.92$). The one-dimensional
observable therefore ranks the cross-section almost as well as the full hybrid
network does, which on this task is barely at all, and placing the encoder's
features beside it recovers the no-circuit level rather than exceeding it.

This pair is also the one place in the design where training the circuit
could show. In the four hybrid conditions the head receives the encoder output
beside the circuit's, so a frozen circuit that already computes a bounded
nonlinear function of that output leaves little for its parameters to add, and
trained-equals-frozen there may be close to structural.
With the encoder withheld from the head, the circuit's parameters are the only
route from features to score. Holding them at their initialisation gives a
pooled Sharpe ratio of \num{-0.183} (95\% CI $[-1.42, +0.91]$, $p = 0.74$)
and an information coefficient of \num{0.0072} ($t = 0.88$), against
\num{-0.124} and \num{0.0075} with them trained. The trained circuit moves by
\SI{20}{\percent} of its initial norm, and the two are no longer
bit-for-bit alike: per-cell Sharpe ratios correlate at \num{0.98} rather than
\num{1.000}, and information coefficients differ by up to \num{0.005} in a
cell. The paired difference is \num{+0.059} Sharpe in the trained circuit's
favour, with a standard error of \num{0.054}, so two one-sided tests certify
equivalence inside a margin of \num{0.25} ($p < 0.001$) and not inside
\num{0.10} ($p = 0.22$); the minimum detectable effect is \num{0.15}. Where
training the circuit is the only thing that can matter, it does not help by
more than a quarter of a Sharpe unit, and the design cannot resolve this
contrast as finely as the hybrid one, which is why the \num{0.03} certificate
belongs to the hybrid contrast alone.

The gradient-boosted baseline does not change the picture. LightGBM on the
flattened window reaches a pooled Sharpe ratio of \num{-0.42} (95\% CI
$[-1.72, +0.78]$, $p = 0.51$), an information coefficient of \num{0.0051}
($t = 1.01$) and a turnover of \SI{72}{\percent} per leg per day, the highest
of the eleven conditions; early stopping keeps \num{14} trees on average. It sits
\num{0.41} Sharpe below the no-circuit control ($p = 0.22$; Diebold--Mariano
$p = 0.44$). Whatever signal these features carry, the model family that wins
this competition on them extracts no more of it than the networks do.

\begin{figure}[t]
\centering
\includegraphics[width=0.55\linewidth]{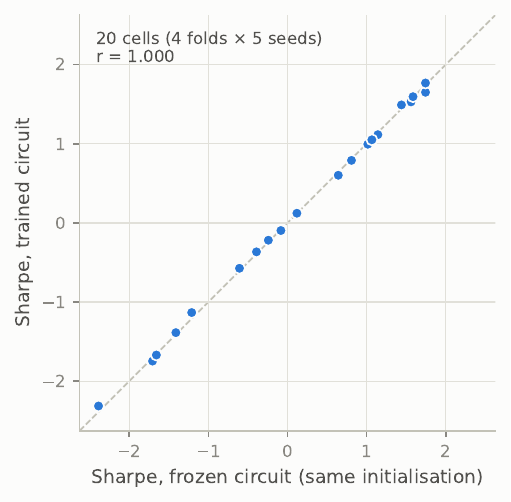}
\caption{Per-cell Sharpe ratio, trained circuit against frozen circuit. The
dashed line is the identity.}
\label{fig:trained-frozen}
\end{figure}

Figure~\ref{fig:trained-frozen} shows the one contrast this design resolves
finely. Because a non-significant difference is not evidence of no difference,
it is tested as an equivalence rather than as a failure to reject. On the pooled track the trained circuit and the same circuit frozen at
its initialisation differ by \num{0.008} in annualised Sharpe ratio, with a
paired stationary-bootstrap standard error of \num{0.0075}. Two one-sided
tests against a margin of \num{0.03} Sharpe give $p = \num{0.0055}$: the
difference is inside that margin, not merely indistinguishable from zero. The
design could have detected a true difference of \num{0.02} at \SI{80}{\percent}
power, so the margin is not an artefact of a blunt instrument. Across the
twenty (fold, seed) cells the two agree to a correlation of \num{1.000}, their
information coefficients never differ by more than $7.7\times10^{-5}$, and their
validation AUCs by more than $2.0\times10^{-4}$.

The contrasts against the no-circuit control are weaker and we report them as
such. Because the study ranks by Sharpe ratio, these are tested twice: by a
Diebold--Mariano statistic on the difference of daily returns, which is a
hypothesis about means, and by a paired bootstrap on the difference of Sharpe
ratios, which is the hypothesis the table is ordered on. The two agree here,
with no contrast significant under either; the closest is the surrogate, \num{0.22}
Sharpe below the control ($p = 0.06$ paired bootstrap, \num{0.09}
Diebold--Mariano), whose paired interval, $[-0.47, -0.0003]$, ends just short of
zero. The circuit and the surrogate are not distinguishable from each other
either: the frozen circuit is \num{0.13} Sharpe above the surrogate on this
track ($p = 0.17$), \num{0.03} above it on the US panel ($p = 0.86$) and
\num{0.06} below it on the holdout ($p = 0.89$), and they are equivalent inside
\num{0.50} on both walk-forward panels. The trained circuit beats that control by \num{-0.082} Sharpe with a
standard error of \num{0.120}; equivalence holds at a margin of \num{0.50}
($p < \num{0.001}$) but not at \num{0.25} ($p = \num{0.07}$), and the minimum
detectable effect at \SI{80}{\percent} power is \num{0.34}. The design
therefore shows that the circuit contributes less than half a Sharpe unit, not
that it contributes nothing. The narrower claim it does establish is that
optimising the circuit's eighteen parameters changes
nothing that freezing them does not change identically, so on this task the
quantum layer acts as a fixed random feature.

The equivalence is not a failure to train: the circuit's parameters move.
Across the twenty walk-forward cells the restored weights sit
\SI{18}{\percent} of the initial norm from where they started (range
\numrange{3}{54}\,\si{\percent}), and \SI{67}{\percent} under the full
training budget below. They move, and nothing downstream registers it: the
validation AUC of the trained and frozen conditions differ by at most
$2.0\times10^{-4}$. Whatever the circuit learns, the encoder and head absorb. This is
the behaviour \citet{bermejo2024quantum} predict for a randomly initialised
quantum convolutional network, which they show acts only on low-bodyness
information about its input and is reproduced by a classical surrogate built
from Pauli propagation. Their demonstration is on physics benchmarks; ours is
the same prediction measured on a financial ranking task. Here it is
corroborated by the controls: neither the circuit nor a parameter-matched
classical surrogate improves on the network with no circuit, and the circuit
and the surrogate are indistinguishable from each other.

Nor is it an artefact of the training budget. Early stopping halts the
circuit conditions after \num{9.1} epochs on average and restores the
best-validation epoch, so the circuit might never have been given time to
matter. Re-running both circuit conditions for the full \num{30} epochs
and scoring the last epoch's weights moves the eighteen parameters much
further, by \SI{67}{\percent} of their initial norm on average over the twenty
cells (range \numrange{25}{140}\,\si{\percent}), and changes almost nothing
downstream: information coefficients differ by at most $3.8\times10^{-4}$ in a
cell, per-cell Sharpe ratios correlate at \num{0.987}, and the pooled
difference is \num{0.000} with a paired standard error of \num{0.043}. That
standard error is six times the early-stopped one, because twenty more epochs
let the encoder and head of the two conditions drift apart, so the
equivalence this control certifies is inside \num{0.10} ($p = 0.02$, minimum
detectable effect \num{0.12}) and not inside \num{0.03} ($p = 0.24$). The
longer budget lowers both conditions equally, from \num{-0.09} to \num{-0.11}
Sharpe, and halves their information coefficient, from \num{0.0083} to
\num{0.0038}: a mild overfitting the pair shares rather than a contribution of
the circuit.

The ordering survives a change of ansatz; the precision does not, and we
report both (Table~\ref{tab:reup}). On the re-uploading circuit with a ring of
entanglers and an observable averaged over all eight wires (Section~\ref{sec:method:arch}), the
trained and frozen conditions again sit below the no-circuit control, by
\num{0.22} and \num{0.18} Sharpe ($p = 0.09$ and \num{0.21}; Diebold--Mariano
$p = 0.08$ and \num{0.26}), their validation AUCs agree to $1.5\times10^{-4}$, their
per-cell Sharpe ratios correlate at \num{0.998}, and training moves the
\num{48} parameters by \SI{21}{\percent} of their initial norm. Two things
differ from the QCNN. The two conditions are no longer bit-for-bit alike:
information coefficients differ by up to \num{0.011} in a cell, where the QCNN
pair differed by at most $7.7\times10^{-5}$, so the all-wire observable does let the
parameters reach the score. And the paired standard error of the pooled
Sharpe difference is \num{0.091}, twelve times the QCNN's, so two one-sided
tests certify equivalence only inside a margin of \num{0.25} ($p = 0.005$;
minimum detectable effect \num{0.25}) and not inside \num{0.03}
($p = 0.52$). The difference itself is \num{-0.032}, in the frozen circuit's
favour. What transfers across the two families is therefore the claim that
optimising the circuit does not improve on freezing it; the claim that it
changes nothing at all, to \num{0.03} Sharpe, is established for the QCNN
and bounded at \num{0.25} for the second family.

\begin{table}[t]
  \centering
  \caption{Second ansatz family (data re-uploading, CNOT ring, mean of $Z$
  over all wires) on the Tokyo walk-forward track, \num{500} days, five
  seeds; the no-circuit control is the one of Table~\ref{tab:headline}.
  Bootstrap 95\% intervals and $p$ against zero; information coefficient with
  its Newey--West $t$; $\Delta$ is the pooled Sharpe difference from the row
  above it with its paired standard error; TOST gives the smallest of the
  margins \num{0.03}, \num{0.10}, \num{0.25} at which equivalence holds at
  \SI{5}{\percent}.}
  \label{tab:reup}
  \small
  \begin{tabular}{lrcrrrr}
    \toprule
    Condition & Sharpe & 95\% CI & $p$ & IC ($t$) & $\Delta$ (s.e.) & TOST \\
    \midrule
    No circuit     & $-0.006$ & [$-1.21$, $+1.13$] & 0.99 & 0.0082 (0.99) & -- & -- \\
    Frozen circuit & $-0.190$ & [$-1.48$, $+0.97$] & 0.75 & 0.0084 (1.04) & $-0.184$ (0.160) & -- \\
    Trained circuit& $-0.222$ & [$-1.48$, $+0.95$] & 0.71 & 0.0077 (0.97) & $-0.032$ (0.091) & 0.25 \\
    \bottomrule
  \end{tabular}
\end{table}

\subsection{Ranking skill}
\label{sec:results:ic}

Sharpe ratios are the metric this study is ordered on, but they confound a
model's ranking skill with the portfolio rule applied to it, and the nearest
prior work on the equity cross-section states its headline contrast as a
difference in information coefficient \citep{shen2026quantum}. We therefore
report the ranking statistics directly (Table~\ref{tab:ic}).

\begin{table}[t]
  \centering
  \caption{Ranking skill on both panels: the mean daily information
  coefficient, its Newey--West $t$-statistic, and the Spearman correlation
  between score decile and realised mean forward return as a monotonicity
  measure. Means over the twenty (JPX) and twenty-five (US) cells per condition.}
  \label{tab:ic}
  \small
  \begin{tabular}{lrrrrrr}
    \toprule
    & \multicolumn{3}{c}{JPX} & \multicolumn{3}{c}{US} \\
    \cmidrule(lr){2-4}\cmidrule(lr){5-7}
    Condition & IC & $t_{\mathrm{IC}}$ & Mono. & IC & $t_{\mathrm{IC}}$ & Mono. \\
    \midrule
    QTCNN, trained circuit     & 0.0083 &  1.01 &  0.054 & 0.0033 &  0.30 & -0.376 \\
    QTCNN, frozen circuit      & 0.0083 &  1.01 &  0.048 & 0.0033 &  0.30 & -0.379 \\
    QTCNN, classical surrogate & 0.0073 &  0.91 &  0.080 & 0.0023 &  0.24 & -0.432 \\
    QTCNN, no circuit          & 0.0082 &  0.99 &  0.114 & 0.0033 &  0.32 & -0.336 \\
    QTCNN, circuit only        & 0.0075 &  0.91 &  0.115 &   --   &   --  &   --   \\
    QTCNN, circuit only, frozen & 0.0072 &  0.88 &  0.119 &   --   &   --  &   --   \\
    LightGBM baseline          & 0.0051 &  1.01 &  0.032 & 0.0034 &  0.50 & -0.149 \\
    TCN baseline               & 0.0059 &  0.83 &  0.049 & 0.0006 &  0.02 & -0.421 \\
    LSTM baseline              & 0.0032 &  0.60 & -0.065 & 0.0052 &  0.46 & -0.163 \\
    Transformer baseline       & 0.0072 &  0.89 &  0.007 & 0.0072 &  0.71 & -0.099 \\
    MLP baseline               & 0.0057 &  0.86 &  0.035 & 0.0009 &  0.06 & -0.318 \\
    \bottomrule
  \end{tabular}
\end{table}

Three observations follow, and the first bears on the ablation. The
trained and frozen circuits agree on the ranking statistics: identical
information coefficients to four decimal places and $t$-statistics to two;
decile monotonicity, which rests on ten portfolio means per cell, differs by
\num{0.005} on JPX and \num{0.003} on the US panel. The equivalence of
Section~\ref{sec:results:ablation} is therefore not an artefact of the
portfolio rule washing out a difference in the scores; the scores themselves
are the same.

Second, the effect sizes place this study where it belongs in the literature.
The largest difference in information coefficient between a condition with a
circuit or surrogate beside the encoder and the no-circuit control is at most
\num{0.001} on either panel, against the
\num{0.005} that \citet{shen2026quantum} report as an indistinguishable
kernel-swap contrast on Chinese A-shares, and no $t$-statistic on either panel
exceeds \num{1.1}. The effects are small, not large and imprecisely
measured.

Third, ranking skill does not translate into realised Sharpe, and on the US
panel it is actively misleading. Every condition has a positive mean
information coefficient there while every condition loses money, and every
decile monotonicity is negative: the score orders the cross-section in the
direction opposite to realised returns at the extremes, which is where a
long--short book trades. The same dissociation, in the other direction, is what
Section~\ref{sec:results:reeval} finds in the reference implementation, whose
best-performing model has the worst information coefficient of nine. A paper
reporting only one of these two families of statistic can reach either
conclusion it prefers.

\subsection{What does vary}
\label{sec:results:seeds}

\begin{figure}[t]
\centering
\includegraphics[width=0.9\linewidth]{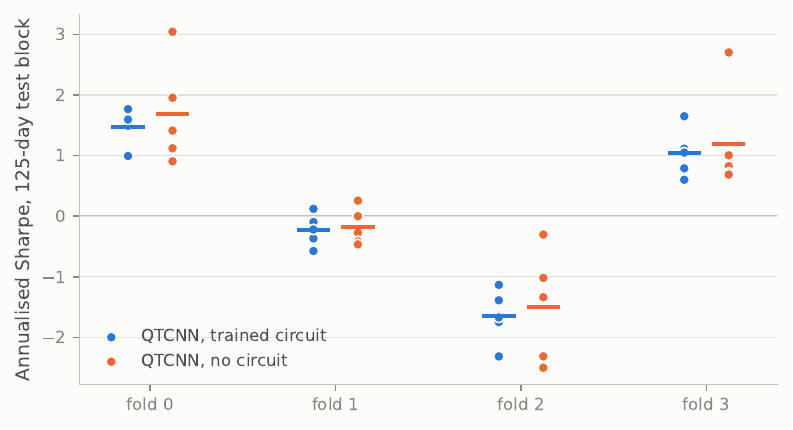}
\caption{Sharpe ratio per 125-day test block, one point per seed, for the
trained-circuit model and the no-circuit control. Bars mark seed means.}
\label{fig:seeds}
\end{figure}

Figure~\ref{fig:seeds} shows why single-seed comparisons of these models cannot
be trusted. The seed controls only which \num{60000} of roughly \num{190000}
labelled training sequences a fold draws, yet within a fold the spread across
seeds (standard deviation \numrange{0.15}{1.70} in Sharpe units, and up to
\num{4.39} between the best and worst seed of one condition) is as large as or
larger than the differences between conditions in Table~\ref{tab:headline},
which reach \num{0.60}. Fold 2, which covers
the 2020 recovery, is negative in \num{51} of its \num{55} cells; fold 3 is
positive in \num{52} of its \num{55}. On this task the training draw and the test regime
matter, and the architecture does not.

The combinatorially symmetric cross-validation estimate of the probability of
backtest overfitting across the eleven conditions is \num{0.58}: choosing the
in-sample best of them lands below the out-of-sample median more often than
not, worse than choosing at random.

\subsection{Transaction costs}
\label{sec:results:costs}

\begin{figure}[t]
\centering
\includegraphics[width=0.75\linewidth]{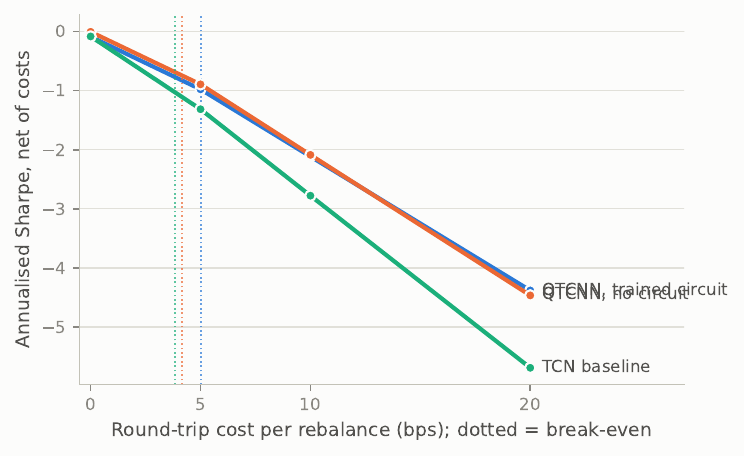}
\caption{Annualised Sharpe net of round-trip transaction costs. Net Sharpe is
linear in cost at fixed turnover; dotted verticals mark break-even.}
\label{fig:cost}
\end{figure}

Turnover runs from \SI{34}{\percent} (LSTM) to \SI{72}{\percent} (LightGBM) of
each leg per day, from \SI{44}{\percent} to \SI{47}{\percent} across the
four hybrid conditions, and \SI{48}{\percent} for the two circuit-only
conditions. Figure~\ref{fig:cost} charges a round-trip cost on the
traded fraction of both legs. On the pooled track, break-even exists for
exactly one of the eleven conditions, the transformer baseline at
\SI{0.49}{bps}, because the other ten have a negative gross mean return
before any cost is charged. Averaging the per-cell break-evens instead, as is
sometimes done, gives a far more flattering \numrange{1.6}{5.0}\,bps, but that
average is taken over only the nine to twelve cells in twenty whose gross
return was positive, and it is not comparable across conditions. At
\SI{5}{bps} the mean per-cell net Sharpe ratio runs from \num{-0.89} to
\num{-2.94}, and at \SI{10}{bps} from \num{-1.85} to \num{-5.63}. The reference implementation's own output, re-scored with the same procedure,
breaks even at \SI{7.4}{bps} on a turnover of \SI{24}{\percent}
(Section~\ref{sec:results:reeval}); neither figure describes a strategy that
could be traded.

\subsection{Is the quantum signal at least orthogonal?}
\label{sec:results:spanning}

A model that loses head-to-head can still earn a place in an ensemble if its
return is not spanned by the alternatives. It is not. Regressing the
trained-circuit model's daily return on the four neural baselines gives
$R^2 = 0.910$ with an annualised intercept of \num{-0.001} and
$t = -0.04$ ($p = 0.97$); regressing it on its own no-circuit twin alone gives
$R^2 = 0.975$ and $t = -0.73$. Pairwise return correlations with the
four baselines run from \num{0.88} to \num{0.94}, and no equal-weight blend
with any of them has a positive Sharpe ratio.

\subsection{A second market}
\label{sec:results:us}

If the null is a property of the task rather than of one exchange, it should
reappear elsewhere. We repeat the entire protocol on the US panel of
Section~\ref{sec:data-us}: the same conditions (nine; the two circuit-only
conditions were run on Tokyo only), the same code path, the
same five seeds, five walk-forward folds giving \num{625} pooled out-of-sample
days. Table~\ref{tab:headline-us} reports the result.

\input{tables/us/headline}

The equivalence replicates. The trained and
frozen circuits differ by \num{-0.002} Sharpe with a paired standard error of
\num{0.0073}; two one-sided tests against a \num{0.03} margin give
$p < \num{0.001}$, and the minimum detectable effect at \SI{80}{\percent}
power is \num{0.02}. Their per-cell Sharpe ratios correlate at \num{0.9994}
and their pooled daily returns at \num{0.9999}. Against the no-circuit
control the picture is again weaker: equivalence holds at a \num{0.50} margin
and not at \num{0.25}, with a minimum detectable effect of \num{0.65}. No
Diebold--Mariano contrast involving a hybrid condition is significant, nor is
any paired bootstrap test of the Sharpe difference itself; the one uncorrected
rejection on this panel is the LightGBM baseline against the control
($p = 0.03$), and nothing survives the Romano--Wolf correction, whose smallest
adjusted $p$ is \num{0.22}, for that same contrast.

What changes is the sign and, unlike the walk-forward on Tokyo data, the level.
Every condition loses money over this window, which runs from March 2019 to
September 2021 and contains both the COVID crash and the 2021 reversal, and for
five of them (both circuit conditions, the surrogate, the TCN and the MLP)
the loss is itself significant at the \SI{5}{\percent} level ($p$ from \num{0.009} to
\num{0.046}); the no-circuit control's interval also excludes zero, at
$p = 0.052$. The gross mean
return is negative for all nine, so no condition has a break-even transaction
cost at all. The four hybrid conditions cluster between \num{-1.44} and
\num{-1.49}, while the classical baselines spread from \num{-0.66} (LSTM) to
\num{-2.03} (TCN) on both sides of them. The LightGBM-versus-control
Diebold--Mariano contrast reaches $p = 0.03$ uncorrected and \num{0.22} after
Romano--Wolf adjustment; no other contrast is significant under any test we
apply. The backtest-overfitting probability is \num{0.15}.

Our first run on this panel was wrong in one respect.
The pairwise return correlations across the eight network conditions run from
\num{0.62} to \num{1.000}, not the near-unity we initially measured: that
figure came from a single contaminated security-day, described in
Section~\ref{sec:data-us}, and the inference we had drawn from it, that the
conditions move as one because they track a single signal, was an artefact
of the error rather than a finding. The equivalence between trained and frozen
survived the correction unchanged, because a shared bad day cancels in a paired
difference; every level statistic on the panel did not.

Taken with Section~\ref{sec:results:ablation}: in two markets, on different
calendars, with different signs of gross performance, optimising the circuit
changes nothing that freezing it does not change identically, to within a
margin of \num{0.03} Sharpe that the design is powered to resolve.

\subsection{Re-evaluating the reference implementation}
\label{sec:results:reeval}

Contribution~4 is an audit, and an audit that only asserts is worth little. We
therefore re-scored the reference implementation's own saved predictions with
the procedure of Section~\ref{sec:protocol}, over the longest out-of-sample
window its outputs support (\num{241} trading days rather than the \num{22}
it reported), leaving its models, features and training untouched.
Table~\ref{tab:reeval} is the result.

\begin{table}[t]
  \centering
  \caption{The reference implementation's nine models, re-scored on
  \num{241} out-of-sample days (2020-12-10 to 2021-12-03) with annualised
  metrics, \SI{95}{\percent} stationary-bootstrap intervals, Deflated Sharpe
  Ratios and Newey--West information-coefficient $t$-statistics. Break-even is
  omitted where the gross return is not positive. Nothing else about the
  pipeline was changed.}
  \label{tab:reeval}
  \small
  \begin{tabular}{lrcrrrrr}
    \toprule
    Model & Sharpe & 95\% CI & $p$ & DSR & IC & $t_{\mathrm{IC}}$ & B/E (bps) \\
    \midrule
    QTCNN (reference)  &  1.80 & [-0.03,  3.79] & 0.07 & 0.54 & 0.0016 &  0.42 & 7.4 \\
    Transformer        &  0.46 & [-0.88,  2.04] & 0.53 & 0.11 & 0.0025 &  0.65 & 2.7 \\
    QLSTM              &  0.13 & [-1.42,  1.75] & 0.85 & 0.06 & 0.0132 &  2.37 & 0.8 \\
    LightGBM ranker    & -0.32 & [-2.34,  1.51] & 0.78 & 0.02 & 0.0060 &  1.52 & --  \\
    QNN                & -0.43 & [-2.31,  1.42] & 0.66 & 0.02 & 0.0148 &  2.10 & --  \\
    MLP                & -0.56 & [-2.24,  1.39] & 0.60 & 0.01 & 0.0134 &  2.57 & --  \\
    LSTM               & -0.59 & [-2.31,  1.27] & 0.49 & 0.01 & 0.0030 &  1.05 & --  \\
    PQC + LightGBM     & -0.76 & [-2.38,  1.06] & 0.39 & 0.01 & 0.0064 &  2.21 & --  \\
    QCNN-LSTM          & -2.15 & [-4.22, -0.17] & 0.04 & 0.00 & 0.0018 &  0.70 & --  \\
    \bottomrule
  \end{tabular}
\end{table}

Three things follow, and none of them requires disagreeing with the reference
implementation about anything except how it was scored.

First, the headline survives the longer window in point estimate but not in
inference. The quantum model is still ranked first, at an annualised Sharpe
ratio of \num{1.80}, but its bootstrap interval spans zero, its $p$-value is
\num{0.07}, and its Deflated Sharpe Ratio is \num{0.54} against the \num{0.95}
a claim requires. Eleven-fold more out-of-sample data does not convert the
result into a significant one; it reveals that the original window was too
short to tell.

Second, and more diagnostically, the ranking by Sharpe ratio and the ranking by
information coefficient are close to opposed. The quantum model has the
\emph{lowest} information coefficient of the nine (\num{0.0016},
$t = \num{0.42}$), while the three models whose information coefficients are
individually significant (MLP at $t = \num{2.57}$, QLSTM at \num{2.37},
PQC\,+\,LightGBM at \num{2.21}, QNN at \num{2.10}) earn little or nothing:
three have negative Sharpe ratios and QLSTM \num{+0.13} ($p = 0.85$). A model that
ranks the cross-section better than its rivals and earns less from it is not
being rewarded for its ranking. On a \num{22}-day window that pattern is
invisible; over \num{241} days it is the most informative thing in the table.

Third, costs decide the rest. The reference model turns over \SI{24}{\percent}
of each leg per day and breaks even at \SI{7.4}{bps}; three of the nine have a
break-even at all, and the best of those is under \SI{8}{bps}. None of the nine
describes a strategy that could be traded at institutional costs, which is a
statement about the task and not about any of the architectures in it.

\subsection{Circuit-level diagnostics}
\label{sec:results:diagnostics}

\begin{figure}[t]
\centering
\includegraphics[width=\linewidth]{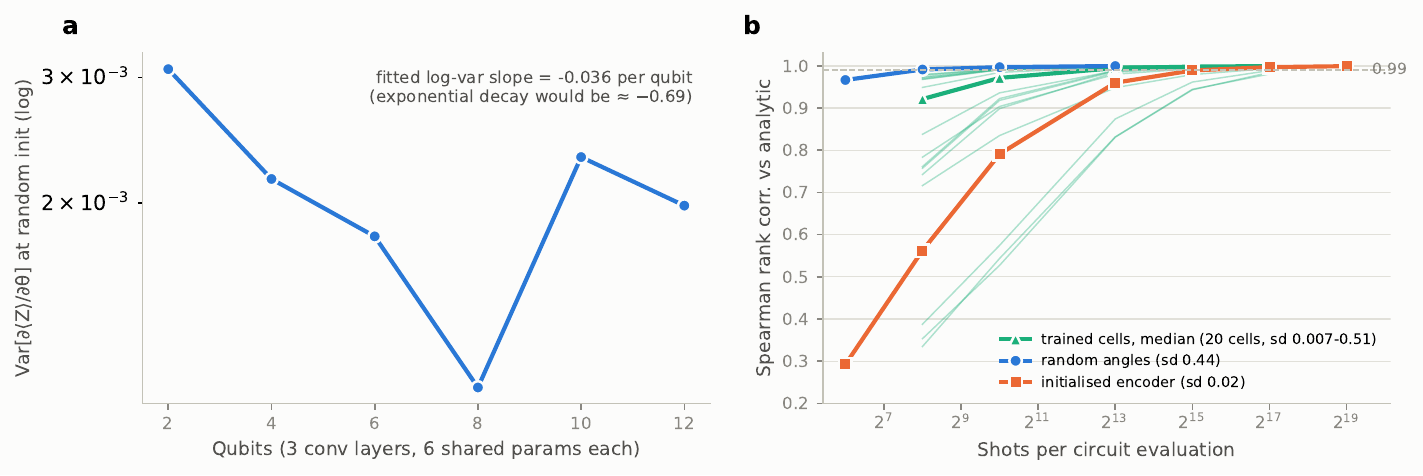}
\caption{Circuit-level diagnostics at eight qubits and three layers.
(a) Gradient variance against register width. (b) Rank preservation against
shot budget at three input distributions; faint lines are the twenty trained
cells.}
\label{fig:diagnostics}
\end{figure}

The negative result above is not explained by the circuit being untrainable.
Figure~\ref{fig:diagnostics}a plots the variance of a circuit parameter's gradient at
random initialisation from two to twelve qubits: it moves from
$3.1\times10^{-3}$ to $2.0\times10^{-3}$, a fitted log-variance slope of \num{-0.036}
per qubit against the \num{-0.69} an exponential $2^{-n}$ decay would give.
The shared-parameter construction keeps the parameter count at $6L$ regardless
of width, and the measured observable is local, which are the conditions under
which barren plateaus are avoided. Expressibility, measured as KL divergence
from the Haar fidelity distribution, falls from \num{0.188} to \num{0.078}
between one and three layers at eight qubits, so the circuit is neither
degenerate nor flat. It is trainable; training it does not help.

Figure~\ref{fig:diagnostics}b addresses deployment, and it is the one diagnostic in
this section whose answer depends on where the circuit's inputs come from. The
strategy consumes a ranking, so the relevant quantity is the Spearman
correlation between finite-shot and analytic scores within a trading date. On
the uniform random angles a circuit-level scan naturally uses, that correlation
is \num{0.991} at \num{256} shots and \num{0.9996} at \num{8192}, and the
implied budget is undemanding.

Those inputs are not the ones the model produces. A first measurement at
the model's input distribution, reported in an earlier version of this
manuscript, pushed real feature windows through an \emph{initialised} encoder
and found a within-date spread of the circuit output of \num{0.018}, a factor
of \num{24} below random angles, and a budget of about \num{32768} shots to
reach a rank correlation of \num{0.99}; the text described that encoder as
trained, which it was not. Repeating the measurement on the twenty trained
cells, each on three dates drawn from its own test window with \num{300}
securities per date, gives a less tidy picture (Table~\ref{tab:shots}).
Training widens the band: the median cell has a within-date standard
deviation of \num{0.10}, and the range across cells runs from \num{0.007} to
\num{0.51}, so the shot budget is a property of the trained cell rather than
of the circuit. Eight of the twenty cells reach a rank correlation of
\num{0.99} at \num{1024} shots, three at \num{8192}, five at \num{32768} and
one at \num{131072}; the three narrowest, all from one seed, do not reach it at
\num{131072}, and no cell reaches it at \num{256}. The median cell therefore
needs \num{8192} shots. Averaged over cells the figure is \num{0.99} at about
\num{32768} shots, which coincides with the initialised-encoder number by
accident, as the mean of cells that need \num{1024} and cells that need more
than \num{131072}. Scoring \num{2000} securities for one trading day
therefore costs about $1.6\times10^{7}$ circuit executions at the median cell
and $7\times10^{7}$ at the mean, against the $5\times10^{5}$ a random-angle
scan implies.

\begin{table}[t]
  \centering
  \caption{Rank preservation against shot budget at three input
  distributions: uniform random angles; real feature windows through an
  initialised encoder (12 dates up to \num{8192} shots and 6 further dates
  above, 300 securities each); and real windows
  through each of the twenty trained cells on dates from its own test window
  (3 dates per cell, 300 securities each). Entries are within-date Spearman
  correlations between finite-shot and analytic scores; for the trained
  cells the median over cells and the range across cells are given.}
  \label{tab:shots}
  \small
  \begin{tabular}{rrrrr}
    \toprule
    & Random angles & Initialised encoder & \multicolumn{2}{c}{Trained cells} \\
    \cmidrule(lr){4-5}
    Shots per security & (sd \num{0.443}) & (sd \num{0.018}) & median (sd \num{0.10}) & range \\
    \midrule
    256     & 0.991  & 0.56  & 0.921 & 0.33--0.990 \\
    1\,024  & 0.997  & 0.79  & 0.972 & 0.53--0.997 \\
    8\,192  & 0.9996 & 0.96  & 0.995 & 0.83--0.9995 \\
    32\,768 & --     & 0.989 & 0.998 & 0.94--0.9998 \\
    131\,072& --     & 0.997 & 0.9995 & 0.98--0.9999 \\
    \bottomrule
  \end{tabular}
\end{table}

The mechanism is a property of the task and of the training trajectory
rather than of the device. Cross-sectional normalisation and the encoder map
a day's securities into a band of embedding angles whose width training sets;
where the band is narrow the observable is concentrated, ranking within it is
a fine discrimination, and finite sampling resolves it slowly. Exponential
concentration of quantum observables is a known obstacle for kernels and
reservoirs as system size grows~\citep{miroszewski2024search}; what this
measurement shows is a task-induced version of it at eight qubits, where the
system is far too small for the asymptotic argument to bite, and one whose
severity differs by two orders of magnitude between cells that are otherwise
statistically indistinguishable. A circuit-level scan on random inputs cannot
see it, and neither can a single measurement on one encoder, which is a reason
to measure shot budgets on the deployed models themselves.

Gate noise, by contrast, is invisible at those budgets. The scan applies a
depolarising channel to every active wire after each convolution layer, so it
is not a single terminal channel and need not act as a pure rescaling of the
readout; in practice its effect on the ordering is negligible, with a rank
correlation above \num{0.9999} at every rate up to $p = 0.05$
whether the rate is uniform across wires or drawn per wire. Under heterogeneous
depolarising noise at \num{1024} shots the correlation is \num{0.997} at
$p = 0$ and \num{0.997} at $p = 0.05$: sampling accounts for all of the
degradation. Amplitude damping, which is non-unital, produces a mean shift in
$\langle Z\rangle$ growing linearly to \num{0.026} at $\gamma = 0.05$ while
depolarising produces none, which confirms the measurement can distinguish the
two, and still leaves the rank correlation above \num{0.9997}. For this
circuit and observable the deployment constraint is the sampling budget, not
device fidelity.

\input{sections/06_results_holdout}

%% file: tables/headline.tex
\begin{table}[t]
\centering
\caption{Out-of-sample performance on the pooled walk-forward track (\num{500} trading days, five seeds per condition). Confidence intervals are 95\% stationary bootstrap; DSR is the Deflated Sharpe Ratio deflated by the \num{55} configurations evaluated.}
\label{tab:headline}
\small
\begin{tabular}{lrrrrrr}
\toprule
Condition & Sharpe & 95\% CI & $p$ & DSR & Seed sd & Max DD \\
\midrule
Transformer baseline & +0.107 & [-1.09, +1.27] & 0.85 & 0.17 & 0.31 & -0.128 \\
QTCNN, no circuit & -0.006 & [-1.21, +1.13] & 0.99 & 0.14 & 0.20 & -0.164 \\
TCN baseline & -0.085 & [-1.28, +1.11] & 0.88 & 0.12 & 0.47 & -0.163 \\
QTCNN, trained circuit & -0.088 & [-1.27, +1.04] & 0.87 & 0.11 & 0.14 & -0.176 \\
QTCNN, frozen circuit & -0.096 & [-1.29, +1.03] & 0.86 & 0.11 & 0.15 & -0.178 \\
QTCNN, circuit only & -0.124 & [-1.34, +0.95] & 0.82 & 0.11 & 0.22 & -0.148 \\
MLP baseline & -0.176 & [-1.34, +0.96] & 0.74 & 0.09 & 0.61 & -0.116 \\
QTCNN, circuit only, frozen & -0.183 & [-1.42, +0.91] & 0.74 & 0.09 & 0.26 & -0.159 \\
QTCNN, classical surrogate & -0.224 & [-1.49, +0.97] & 0.71 & 0.08 & 0.35 & -0.192 \\
LightGBM baseline & -0.419 & [-1.72, +0.78] & 0.51 & 0.05 & 0.38 & -0.135 \\
LSTM baseline & -0.496 & [-1.69, +0.81] & 0.44 & 0.04 & 0.13 & -0.233 \\
\bottomrule
\end{tabular}
\end{table}

%% file: tables/us/headline.tex
\begin{table}[t]
\centering
\caption{Out-of-sample performance on the pooled walk-forward track for the US panel (\num{625} trading days, five seeds per condition). Confidence intervals are 95\% stationary bootstrap; DSR is the Deflated Sharpe Ratio deflated by the \num{55} configurations evaluated.}
\label{tab:headline-us}
\small
\begin{tabular}{lrrrrrr}
\toprule
Condition & Sharpe & 95\% CI & $p$ & DSR & Seed sd & Max DD \\
\midrule
LSTM baseline & -0.655 & [-1.83, +0.50] & 0.28 & 0.00 & 0.55 & -0.402 \\
LightGBM baseline & -0.672 & [-2.00, +0.58] & 0.32 & 0.00 & 0.32 & -0.229 \\
Transformer baseline & -1.158 & [-2.34, +0.09] & 0.087 & 0.00 & 0.42 & -0.503 \\
QTCNN, no circuit & -1.441 & [-2.75, -0.05] & 0.052 & 0.00 & 0.16 & -0.532 \\
QTCNN, frozen circuit & -1.463 & [-2.75, -0.09] & 0.046 & 0.00 & 0.23 & -0.535 \\
QTCNN, trained circuit & -1.464 & [-2.74, -0.09] & 0.043 & 0.00 & 0.22 & -0.535 \\
QTCNN, classical surrogate & -1.491 & [-2.79, -0.14] & 0.036 & 0.00 & 0.28 & -0.535 \\
MLP baseline & -1.563 & [-2.92, -0.17] & 0.038 & 0.00 & 0.46 & -0.481 \\
TCN baseline & -2.026 & [-3.38, -0.64] & 0.009 & 0.00 & 0.27 & -0.530 \\
\bottomrule
\end{tabular}
\end{table}

%% file: sections/06_results_holdout.tex
\subsection{Locked holdout}
\label{sec:results:holdout}

The supplemental window, 2021-12-06 to 2022-06-24 (\num{135} trading days),
was evaluated once per condition after that condition's design was frozen,
with each condition trained on all data to 2021-08-06 under the same validation
and embargo rules. It was opened for the eight network conditions, then for
the LightGBM baseline, and the eight network conditions were then re-run on it
under the released code because the first run could not be reproduced
(Section~\ref{sec:protocol:holdout}). Table~\ref{tab:holdout} reports the
reproducible run; Appendix~\ref{sec:app:rerun} gives the first.

\begin{table}[t]
\centering
\caption{Locked holdout, 135 trading days. Uncorrected
bootstrap $p$ against zero; Romano--Wolf $p$ across the eight contrasts with
the no-circuit control on this window (the two circuit-only
conditions and the second ansatz family were not); DSR deflated by the
\num{55} configurations (eleven models, five seeds) the study evaluated, using the
empirical dispersion of their Sharpe ratios on this window (sd \num{0.92},
expected maximum under the null \num{2.12}); break-even computed on the
pooled track, undefined where the gross return is not positive.}
\label{tab:holdout}
\small
\begin{tabular}{lrrrrrr}
\toprule
Condition & Sharpe & Seed sd & $p$ & $p_{\mathrm{RW}}$ & DSR & Break-even (bps) \\
\midrule
QTCNN, no circuit          & $+1.83$ & 0.84 & 0.07 & --   & 0.42 & 3.7 \\
QTCNN, classical surrogate & $+1.57$ & 1.08 & 0.18 & 0.95 & 0.34 & 5.5 \\
QTCNN, frozen circuit      & $+1.51$ & 1.05 & 0.17 & 0.95 & 0.33 & 4.4 \\
QTCNN, trained circuit     & $+1.49$ & 1.05 & 0.18 & 0.96 & 0.32 & 4.4 \\
TCN baseline               & $+1.37$ & 0.77 & 0.22 & 0.95 & 0.29 & 3.0 \\
Transformer baseline       & $+0.68$ & 0.87 & 0.47 & 0.54 & 0.15 & 2.1 \\
LSTM baseline              & $+0.67$ & 0.79 & 0.48 & 0.91 & 0.15 & 3.8 \\
LightGBM baseline          & $+0.49$ & 0.76 & 0.72 & 0.33 & 0.12 & 0.6 \\
MLP baseline               & $+0.42$ & 0.77 & 0.68 & 0.37 & 0.11 & 1.1 \\
\bottomrule
\end{tabular}
\end{table}

The window replicates the ablation and not the sign. Trained and frozen
circuits differ by \num{0.02}; the no-circuit control is the best of the four
hybrid conditions and the surrogate is no worse than either circuit; every
Diebold--Mariano contrast between a hybrid condition and the no-circuit control
has $p > 0.70$ (the classical baselines are closer, at \num{0.14} for LightGBM
and \num{0.16} for the MLP; one paired Sharpe-difference test, the MLP's,
reaches $p = 0.046$ uncorrected; none survives correction). Gross Sharpe
ratios are higher for all nine conditions than on the walk-forward track, which is a property of the
period, a stretch on which the cross-section rewarded the momentum and
liquidity features every model shares. The ranking of the nine is not stable
across the two windows: the transformer moves from first to sixth and the
no-circuit control from second to first, while the seed standard deviations
(\numrange{0.76}{1.08}) are wider than before because the window is short. What
is stable is the ablation itself (trained beside frozen, the surrogate no
worse than either, and no contrast significant), which is the quantity the
design was built to measure.

No condition is distinguishable from zero by the test on this window either.
For the best, the no-circuit control at \num{+1.83}, the percentile interval
just excludes zero ($[+0.02, +3.94]$) while the null-imposed bootstrap gives
$p = 0.07$, and its Deflated Sharpe Ratio is \num{0.42} against the
\num{0.95} a claim would require: deflated by the \num{55} configurations, the
expected maximum Sharpe under the null on \num{135} days is \num{2.12}, above
every row of the table. At \SI{5}{bps} one condition remains positive, the
surrogate at \num{+0.18}; at \SI{10}{bps} none do.

%% file: sections/07_discussion.tex
\section{Discussion}
\label{sec:discussion}

\subsection{What the ablation identifies}
The four hybrid conditions were designed so that their pairwise contrasts each
answer one question, and on this data every answer is the same.
\textsc{Trained} against \textsc{Frozen} asks whether optimising the circuit
matters: the per-cell Sharpe ratios correlate at \num{1.000}, so it does not.
\textsc{Frozen} against \textsc{Surrogate} asks whether the particular random
feature a quantum circuit computes differs from a classical one of equal
parameter count and matching output scale: on no panel can the two be told
apart, and the sign of the difference changes between panels. \textsc{Surrogate}
against \textsc{None} asks whether any extra scalar feature helps the head at
all: it does not. The circuit is a random feature, the random feature is not
special, and the head does not need it.

The same three contrasts give the same three answers on the US panel
(Section~\ref{sec:results:us}), where gross performance is negative rather than
positive. A result that survives a sign change in the underlying market is
about the layer, not about the window.

A null result on a task where nothing works might say little about the layer.
The locked holdout is the positive control for that concern. On those \num{135} days the shared signal is profitable for all
nine conditions, with gross annualised Sharpe ratios from \num{+0.42} (MLP)
to \num{+1.83} (the no-circuit control), and the ablation is unchanged there: trained beside frozen, the
surrogate no worse than either, and no contrast significant. The circuit adds
nothing in the regime where the features fail and nothing in the regime where
they work.

The result does not show that quantum convolutional layers are useless. It
concerns this task: daily cross-sectional ranking from twenty days of
technical features, where the achievable information coefficient is around
\num{0.006} and the spread of one model's Sharpe ratio across seeds
(standard deviation \numrange{0.13}{0.61}) is as large as the spread between
models (\num{0.17}). A layer cannot demonstrate value in a regime where the
signal it would need to improve on is itself indistinguishable from noise.

\subsection{Why earlier numbers looked different}
The reference study this work audits \citep{chen2025quantumb} reports an
out-of-sample Sharpe ratio of \num{0.538} $\pm$ \num{0.042} for its quantum
model, about \SI{72}{\percent} above its best classical baseline, and defines
that ratio as annualised. Its accompanying public implementation, which is the
artefact we re-ran, computes the ratio as mean over standard deviation with no
$\sqrt{252}$ factor and yields \num{0.33} on the same window; on the
conventional scale that is \num{5.2}. Whichever figure one takes, four things,
each individually sufficient, separate it from Table~\ref{tab:headline}. It
was computed on \num{22} trading days, a window over which the bootstrap
interval spans several Sharpe units in either direction; the authors
themselves note that \num{22} days give limited power and that about \num{45}
would be needed for \SI{90}{\percent} power, and defer a full-sample backtest
to future work. The headline is a single-seed model selection (noise
experiments use five runs), and the seed draw alone moves a fold's Sharpe by
up to \num{2.2} units for the two conditions in Figure~\ref{fig:seeds} and by
\num{4.39} for LightGBM. The split adjustment
ran forward in time, corrupting every momentum and volatility feature with
artefactual returns of up to \num{2300}\%. And the paper's own encoder-only
ablation already scores \num{0.412} $\pm$ \num{0.039}, so most of the reported
edge over the Transformer baseline is the temporal encoder, not the circuit, the same decomposition our \textsc{None} condition makes explicit. None of
these is exotic; all are common in the applied quantum machine learning
literature \citep{shen2026quantum,bowles2024better}, where evaluation practice
has not yet caught up with the standards of empirical finance.

\subsection{The circuit is trainable; that is not the problem}
The diagnostics rule out the explanation a quantum-computing reader would reach
for first. Gradient variance is flat from two to twelve qubits, so there is no
barren plateau; expressibility improves with depth, so the ansatz is not
degenerate; the rank of its outputs survives depolarising and amplitude-damping
noise at rates well above current hardware. The layer could be trained, and it
could be run. It has nothing to contribute here that the classical
encoder around it does not already provide.

\subsection{What a deployment would actually cost}
Because the strategy consumes a ranking, the operative hardware question is how
many shots preserve rank order at the spread of the scores being ranked, and
that spread is set by the trained model, not by the circuit. Across the twenty
trained cells the within-date standard deviation of the circuit output runs
from \num{0.007} to \num{0.51}, with a median of \num{0.10}, because
cross-sectional normalisation and the encoder map a day's securities into a
band of angles whose width the training trajectory decides. Preserving the
ranking to a Spearman correlation of \num{0.99} takes about \num{8192} shots
per security at the median cell, roughly $1.6\times10^{7}$ circuit executions
to score \num{2000} securities for one trading day, and more than
\num{131072} at the narrowest cells; the \num{256} a scan on random inputs
suggests is reached by no cell, and the widest need \num{1024}. Gate noise remains negligible at those budgets
(sampling accounts for essentially all of the degradation), so the cost is set
by measurement statistics alone.

We make no claim about whether a present device sustains that rate, and the
comparison that matters is not with a device specification but with the
classical alternative: the same ranking is produced by a parameter-matched
classical map in microseconds, and by no map at all equally well. The
conclusion is therefore narrower than the one a circuit-level scan would
support. It is not that hardware is no obstacle; it is that the obstacle is
mis-stated when the shot budget is measured on random inputs, or on one
encoder, and that on this task the observable can be concentrated enough to
make ranking expensive even at eight qubits, by an amount that varies a
hundredfold between statistically indistinguishable cells. Whether that
concentration is intrinsic to cross-sectional ranking or an artefact of this
encoder is the question we would put first in
any follow-up.

\subsection{Threats to validity}
\label{sec:discussion:threats}
We list the ways this study could be wrong, in decreasing order of how much
they would change the conclusion.

\paragraph{One feature set, chosen by us.}
All eleven conditions see the same \num{21} technical features. If the quantum
layer has an inductive bias suited to structure these features do not contain
(fundamentals, order-flow, cross-asset signals), the null is a statement
about the inputs rather than the layer. Nothing in the design can rule this
out, and it is the most likely way the result fails to generalise.

\paragraph{Two ansatz families, one of them resolved finely.}
The QCNN (angle embedding, shared two-qubit blocks, pooling, single-wire
measurement) and a re-uploading circuit with a CNOT ring and a readout
averaged over every wire give the same ordering, but only the first is certified equivalent
to its frozen copy inside \num{0.03} Sharpe; the second is bounded at
\num{0.25}, because its all-wire readout lets the parameters reach the
score and the paired standard error grows twelvefold. A family in which
training the circuit helps may exist; neither of these two is it, and the
second shows that a looser bound is what a more expressive circuit buys here.

\paragraph{Low achievable signal.}
The information coefficient available on this task is around \num{0.006}
(Table~\ref{tab:ic}), and the spread across seeds is as large as the
differences between conditions. A layer cannot demonstrate value where the signal
it would improve on is itself indistinguishable from noise. The locked holdout
is the partial answer: on a window where the shared signal does pay, the
ablation is unchanged. It is \num{135} days, which is short by this paper's own
standard. The gradient-boosted baseline is the other partial answer: the model
family that wins this competition extracts an information coefficient of
\num{0.005} from the same features, no more than the networks, so the weak
signal is a property of the inputs rather than of the architectures.

\paragraph{Statistical resolution.}
The design resolves the trained-versus-frozen contrast to a margin of
\num{0.03} Sharpe and the contrasts against the no-circuit control only to
\num{0.50}. The second of those is a weak bound and we do not claim more from
it. Four folds and five seeds give twenty paired cells, which is enough for the
first comparison and not for the others.

\paragraph{Simulation only.}
Every circuit result is from a state-vector simulator. The noise scans are
simulated channels, not device runs, and the shot analysis assumes independent
sampling. No claim here has been checked on hardware.

\paragraph{Survivorship in the auxiliary panels.}
The US universe comes from a current exchange symbol directory and excludes
firms delisted before it. A 2022--2025 extension of the JPX panel was also
assembled from the same source for a delisting audit; no result in this paper
uses it, because it is missing
\num{207} of the \num{2000} universe securities, of which the exchange's
official delisting lists account for all: \num{204} left through takeovers,
management buy-outs or squeeze-outs, one failed the listing-maintenance
standards, one was removed by the exchange and one delisted voluntarily. The
JPX development panel itself has the same character: its \num{2000} securities
are the competition's primary universe, fixed by the organiser at the end of
the training period (Section~\ref{sec:data}). All three universes are common
to every condition and to the earlier result we re-score, so the bias enters
no contrast, but absolute performance levels on any of them are not claimed. The US panel also
required a point-in-time price screen to remove a bankruptcy-reorganisation
splice (Section~\ref{sec:data-us}); we found it by auditing for impossible
returns, and a study that did not audit would not have found it.

\paragraph{Fold count.}
Four walk-forward folds are what the JPX sample supports; a fifth would need
\num{44} more trading days than exist. The US calendar affords five.

\subsection{What would change the conclusion}
A positive result would require a task where the circuit's inductive bias
matches structure in the data that the classical encoder cannot express. The
present data has no such structure to find: over the \num{500} development
days only one condition has a positive gross return, and a transaction cost of
half a basis point removes it; on the \num{135}-day holdout the best
break-even cost is \SI{5.5}{bps}. Claims of quantum
advantage in financial prediction should be accompanied by the control
conditions used here (a frozen circuit and a parameter-matched surrogate at
minimum) and by an evaluation long enough to resolve the effect being
claimed. Where they are, we expect most of them to look like
Table~\ref{tab:headline}.

%% file: sections/08_conclusion.tex
\section{Conclusion}
\label{sec:conclusion}

We evaluated a hybrid quantum temporal convolutional network for daily
cross-sectional equity ranking under a protocol built to make the quantum
contribution identifiable: six conditions trained through one code path, four rolling-origin folds with an embargo, five seeds, annualised risk
metrics with bootstrap intervals, Deflated Sharpe Ratios, backtest-overfitting
probabilities and family-wise error control. On \num{500} out-of-sample
trading days the trained circuit is indistinguishable from the same circuit
frozen at initialisation, from a parameter-matched classical surrogate, and
from the network with the circuit removed; no condition is distinguishable from
zero, and ten of the eleven have no break-even transaction cost at all. Two
one-sided tests place the trained-versus-frozen difference inside \num{0.03}
annualised Sharpe in both markets, and inside \num{0.10} when the circuit is
trained to the full budget without early stopping, which is equivalence rather than a failure
to reject; the contrast against the no-circuit control is resolved only to
\num{0.50}.

The circuit is not the weak point. It has no barren plateau to twelve qubits,
and its output ranking is robust to gate noise; a second ansatz family, with
re-uploading and an observable read from every wire, reproduces the ordering and bounds its
trained-versus-frozen difference at \num{0.25} Sharpe. What is
expensive is measurement: at the spread the trained models produce on a real
cross-section, preserving the ranking takes about \num{8192} shots per
security at the median cell, and more than \num{131072} at the narrowest,
rather than the \num{256} a scan on random inputs suggests. The weak point is the task, on which the achievable signal is
smaller than the variance introduced by the training draw. Earlier reports of
advantage on this data came from a 22-day window, single-seed model selection,
a split adjustment applied in the wrong direction in the public implementation,
and a classical encoder that already accounts for most of the reported gain.

Two artefacts are released with the paper \citep{chen2026repo}: a batched formulation of the quantum
convolutional layer that reproduces the per-sample implementation to
double precision at a \numrange{33}{72}-fold speed-up, and the evaluation stack itself,
calibrated on synthetic data with known answers. Both are reusable on any
cross-sectional ranking task. We hope the second is used more often than the
first.

%% file: sections/A_reproducibility.tex
\section{Reproducibility}
\label{sec:repro}

\subsection{Software and hardware}
Experiments ran on a 24-core workstation shared with unrelated jobs, under
PennyLane 0.42.3, PyTorch 2.6.0, NumPy 2.2.6, pandas 2.2.3 and LightGBM 4.7.0 on Python
3.12.9. Quantum circuits were simulated on the \texttt{default.qubit}
state-vector backend with backpropagation. All reported models are trained on
CPU: for an eight-qubit register the simulator is CPU-bound, and a GPU gives no
benefit.

\paragraph{Measurement under contention.}
Wall-clock figures are sensitive to what else the machine is running. The same
training step measured \SI{16.4}{\second} at a one-minute load average of 57 on
24 cores and \SI{0.38}{\second} once the machine was scheduled, a factor of 43.
Every timing reported here was taken through a job scheduler that admits work
against declared CPU, RAM and VRAM, and the load at measurement time is
recorded alongside the figure.

\subsection{Determinism}
Each (model, fold, seed) cell seeds PyTorch, NumPy and the training-set
subsample from the same integer. Circuit parameters are initialised from a
dedicated generator so that the \textsc{Frozen} condition draws exactly the same
weights as the \textsc{Trained} condition's initialisation. Results are
checkpointed per cell, so an interrupted sweep resumes without recomputing
completed work and without changing any seed.

\subsection{Data and code}
The dataset is the public JPX Tokyo Stock Exchange
panel~\citep{jpx2022competition}. The analysis code, including the evaluation
statistics, the walk-forward protocol and the batched circuit, is released with
the paper~\citep{chen2026repo}. The calibration checks reported in
Section~\ref{sec:protocol:stats} (in particular that the
backtest-overfitting estimator returns \num{0.53(4)} on independent random
strategies) are part of the test suite.

\subsection{Deviations from the reference implementation}
The pipeline this work audits (our own, as stated in
Section~\ref{sec:intro}) differs in four respects that materially affect its
numbers, each corrected here: the split adjustment direction
(Section~\ref{sec:data:adjustment}); the Sharpe ratio, which the released code
computes as mean over standard deviation with no annualisation, though the
preprint states that its metric is annualised; the evaluation window,
which date subsampling reduced to \num{22} trading days; and the absence of
transaction costs. Section~\ref{sec:results:reeval} reports the original models
re-scored over \num{241} days with all four corrected, so that the difference
is attributable rather than asserted.

\subsection{The first runs and the reproducible rerun}
\label{sec:app:rerun}
The network results first reported for this study were produced between 6 and
25 September 2026 from working copies of the code that were never committed. When trained weights
were needed for the shot-budget measurement of Section~\ref{sec:results:diagnostics},
re-training a cell with the committed code gave a different model. The
committed code is deterministic: the same cell, trained five times at one
and at four threads, was bit-for-bit identical every time, and the second
ansatz family, first run in October with committed code, reproduces exactly.
The first committed version of the code, from 12 September, gives the same
cell bit-for-bit as the released one, so no committed change altered training.
The September working copies differed from every commit in ways that were not
recorded; the full-budget control of 25 September, for one, scored final-epoch
weights, which no committed version ever did. Every network condition run in September was therefore
re-run with the released code, and the paper reports the rerun.
Table~\ref{tab:rerun} sets the two side by side.

\input{tables/rerun}

The two runs of a condition agree in the aggregate and disagree cell by cell.
On the walk-forward panels per-cell Sharpe ratios correlate at
\numrange{0.40}{0.82} between runs, and pooled Sharpe ratios differ by up to
\num{0.25} on the Tokyo panel, \num{0.73} on the US panel and \num{1.66} on the
\num{135}-day holdout. That is the size of the training-path variation
Section~\ref{sec:results:seeds} documents across seeds, and it is a second
demonstration of the same point: on this task the training draw moves a
condition's Sharpe ratio by more than any architectural choice does. What
depends on the run is therefore the ordering of conditions and the level of any
one of them: the transformer is first on the holdout in one run and sixth in
the other, three US conditions are significantly negative in one and five in
the other, and the single condition with a positive gross walk-forward return
is the no-circuit control in one and the transformer in the other. What does
not depend on it is every contrast the paper rests on. In both runs, on both
markets, the trained and frozen circuits are equivalent inside \num{0.03}
Sharpe, their per-cell Sharpe ratios correlate above \num{0.999}, no condition
on the Tokyo walk-forward is distinguishable from zero, no contrast against the
no-circuit control survives multiplicity correction on any panel, and on the
holdout the two circuit conditions differ by \num{0.02}. The September files are
kept in the repository beside the rerun.

%% file: tables/rerun.tex
\begin{table}[t]
\centering
\caption{The first (September 2026) and the reproducible (November 2026) run of every network condition that was re-run: pooled-track annualised Sharpe ratio of each, and the correlation between the two runs' per-cell Sharpe ratios (not given for the five-cell holdout). The last line of each panel is the trained-minus-frozen pooled Sharpe difference with its paired standard error and the two one-sided test against a 0.03 margin.}
\label{tab:rerun}
\small
\begin{tabular}{lrrr}
\toprule
Condition & September & November & Cell corr. \\
\midrule
\multicolumn{4}{l}{\emph{JPX walk-forward}} \\
\quad trained circuit & $-0.07$ & $-0.09$ & 0.78 \\
\quad frozen circuit & $-0.07$ & $-0.10$ & 0.78 \\
\quad classical surrogate & $-0.04$ & $-0.22$ & 0.79 \\
\quad no circuit & $+0.01$ & $-0.01$ & 0.44 \\
\quad circuit only & $-0.25$ & $-0.12$ & 0.82 \\
\quad TCN baseline & $-0.24$ & $-0.08$ & 0.53 \\
\quad LSTM baseline & $-0.33$ & $-0.50$ & 0.81 \\
\quad Transformer baseline & $-0.14$ & $+0.11$ & 0.55 \\
\quad MLP baseline & $-0.15$ & $-0.18$ & 0.57 \\
\quad trained $-$ frozen (s.e.) & \multicolumn{3}{l}{Sept.\ $-0.001$ (0.0072), $p_{0.03}<0.001$; Nov.\ $+0.008$ (0.0075), $p_{0.03}=0.0055$} \\[2pt]
\multicolumn{4}{l}{\emph{JPX, full budget}} \\
\quad trained circuit & $-0.17$ & $-0.11$ & 0.47 \\
\quad frozen circuit & $-0.17$ & $-0.11$ & 0.51 \\
\quad trained $-$ frozen (s.e.) & \multicolumn{3}{l}{Sept.\ $-0.006$ (0.0090), $p_{0.03}=0.0085$; Nov.\ $+0.000$ (0.0432), $p_{0.03}=0.243$} \\[2pt]
\multicolumn{4}{l}{\emph{US walk-forward}} \\
\quad trained circuit & $-1.44$ & $-1.46$ & 0.71 \\
\quad frozen circuit & $-1.43$ & $-1.46$ & 0.72 \\
\quad classical surrogate & $-1.24$ & $-1.49$ & 0.46 \\
\quad no circuit & $-1.24$ & $-1.44$ & 0.71 \\
\quad TCN baseline & $-1.51$ & $-2.03$ & 0.58 \\
\quad LSTM baseline & $-0.85$ & $-0.66$ & 0.61 \\
\quad Transformer baseline & $-0.84$ & $-1.16$ & 0.40 \\
\quad MLP baseline & $-0.83$ & $-1.56$ & 0.74 \\
\quad trained $-$ frozen (s.e.) & \multicolumn{3}{l}{Sept.\ $-0.005$ (0.0075), $p_{0.03}<0.001$; Nov.\ $-0.002$ (0.0073), $p_{0.03}<0.001$} \\[2pt]
\multicolumn{4}{l}{\emph{Holdout}} \\
\quad trained circuit & $+1.54$ & $+1.49$ & -- \\
\quad frozen circuit & $+1.53$ & $+1.51$ & -- \\
\quad classical surrogate & $+1.70$ & $+1.57$ & -- \\
\quad no circuit & $+1.12$ & $+1.83$ & -- \\
\quad TCN baseline & $+0.72$ & $+1.37$ & -- \\
\quad LSTM baseline & $+1.60$ & $+0.67$ & -- \\
\quad Transformer baseline & $+2.34$ & $+0.68$ & -- \\
\quad MLP baseline & $-0.31$ & $+0.42$ & -- \\
\quad trained $-$ frozen (s.e.) & \multicolumn{3}{l}{Sept.\ $+0.018$ (0.0177), $p_{0.03}=0.255$; Nov.\ $-0.022$ (0.0269), $p_{0.03}=0.357$} \\[2pt]
\bottomrule
\end{tabular}
\end{table}